\documentclass[10pt,tikz,floatfix,showpacs]{revtex4-2}
\usepackage[utf8]{inputenc}
\usepackage{amsthm}
\usepackage{comment}
\usepackage[T1]{fontenc}    
\usepackage[british]{babel}  
\usepackage[sc,osf]{mathpazo}
\usepackage[scaled=0.86]{berasans}  
\usepackage[colorlinks=true, citecolor=blue, urlcolor=blue, linkcolor=blue]{hyperref}  
\usepackage{float}
\usepackage{graphicx} 
\usepackage{subfig}
\usepackage{bitset}
\usepackage[babel]{microtype}  
\usepackage{tikz}
\usetikzlibrary{arrows.meta, positioning,shapes,calc}
\usepackage{amsmath,amssymb,bm,amsfonts,mathrsfs,bbm} 
\usepackage{xspace}  
\usepackage{pgfplots}
\usepackage{xcolor,colortbl}
\def\ba{\begin{equation}}
	\def\ea{\end{equation}}
\def\bea{\begin{eqnarray}}
	\def\eea{\end{eqnarray}}
\def\ben{\begin{equation*}}
	\def\een{\end{equation*}}
\def\bean{\begin{eqnarray*}}
	\def\eean{\end{eqnarray*}}
\def\bma{\begin{mathletters}}
	\def\ema{\end{mathletters}}
\def\bi{\begin{itemize}}
	\def\ei{\end{itemize}}

\newcommand{\be}{\begin{equation}}
	\newcommand{\ee}{\end{equation}}

\newcommand{\ket}[1]{\ensuremath{|#1\rangle}}

\newcommand{\bra}[1]{\ensuremath{\langle#1|}}

\newcommand{\kommentar}[1]{}

\newcommand{\forget}[1]{}

\newtheorem{theorem}{Theorem}
\newtheorem{definition}{Definition}

\newtheorem{corollary}{Corollary}[theorem]

\begin{document}
	
	\title{Network nonlocality breaking channels}
	\author{Kaushiki Mukherjee}
	\email{kaushiki.wbes@gmail.com}
	\affiliation{Department of Mathematics, Government Girls' General Degree College, Ekbalpore, Kolkata-700023, India.}
	
	\author{Nirman Ganguly}
	\email{nirmanganguly@hyderabad.bits-pilani.ac.in, nirmanganguly@gmail.com}
	\affiliation{Department of Mathematics, Birla Institute of Technology and Science, Pilani, Hyderabad Campus, Jawahar Nagar, Kapra Mandal, Medchal District, Telangana 500078, India}
	
	
	\begin{abstract}
	Network nonlocality, a recently noted form of nonlocality has been shown to have distinctive features, marking a significant departure from the notion of standard Bell nonlocality in the context of quantum correlations. On a pragmatic front, it has gained significant importance as researchers worldwide continue to engage in the study on quantum networks. However, as typical to any quantum resource, network nonlocality is also vulnerable to environmental noise, which sometimes prove to be detrimental. Environmental interactions are modeled in terms of quantum channels. In the present study, we introduce and characterize network nonlocality breaking channels. Network nonlocality breaking channels model environmental influences which result in the loss of resource, i.e., the system loses its nonlocal resource due to such interactions. The study is done in the ambit of some suitably chosen inequalities in (i) linear networks and (ii) star-shaped networks. Further, the loss in full network nonlocality is also studied. The work characterizes single qubit unital channels and certain classes of single qubit non-unital channels in this context. Furthermore, we also characterize quantum channels according to their ability in preserving quantum resources, i.e., they do not break network nonlocality, which enables one to identify useful quantum channels in networks. The study is vindicated by illustrations from various noise models like depolarizing and dephasing channels.  	
	\end{abstract}
	\date{\today}
	\maketitle
	
	
	\section{Introduction}\label{intro}

One of the significant aims in the field of quantum information and computation is to study and harness quantum resources. Quantum resource theory \cite{gourresource} studies different resources which are genuinely quantum, in the sense that they can provide quantum advantage over classical conventional schemes \cite{quantadpreskill}. Such quantum resources include superposition \cite{supFeynman}, entanglement \cite{enthoro}, nonlocality \cite{Bel,brunrev} etc. 
	
	Quantum states undergo evolution. In quantum information science, such evolutions are modeled in terms of quantum channels \cite{WildeQIT,WatrousQIT}. Quantum channels are superoperators which take density matrices to density matrices. In mathematical terms, they are trace preserving completely positive maps \cite{WildeQIT,WatrousQIT} (details are given in section \ref{prelim}). An alternative way to state the above is that quantum systems interact with environment and such interactions are characterized by quantum channels. Albeit its abstract formulation, quantum channels provide an important technique to study evolutions of quantum states and has numerous practical applications, one of them being in the domain of quantum error correction \cite{bookerror}. Quantum channels also act as mediums to transmit quantum information, and thus capacity of quantum channels to transmit information is a vibrant area of research \cite{capacitysmith}.  
	
	Environmental interactions can result in the loss of resources. A state which was in possession of a resource may lose it when acted upon by a channel. Usually, such actions are modeled in the following way: Suppose that $\rho$ represents a resourceful quantum state. A channel acts on one part of the state without affecting the other part. Then, the output $\rho^\prime$ loses the resource. For example, when entanglement breaking channels \cite{entbreak} act on one part of a quantum state then the output is always separable. In a similar spirit there are nonlocality breaking channels \cite{nlcbreak}, partially entanglement breaking channels (also known as Schmidt number breaking channels) \cite{SNbreak1,SNbreak2}, negative conditional entropy breaking channels \cite{NCEbreak1,NCEbreak2}. 
	
	Nonlocality is counted as a significant resource, where interests have ranged from its foundational aspects  to practical use \cite{brunrev}. Nonlocal correlations emerging from a single quantum source is detected via violation of Bell's inequality compatible with corresponding measurement scenario. With advancement in technology, study of quantum nonlocality has underwent rapid change. Over the years,  researchers have extensively worked on exploiting nonlocality from quantum network structures which involve more than one source \cite{BGP,BRGP,Tava,Tavak,Tavako}. In this context several key concepts such as source independence($n$-local constraint), $n$-local correlations, $n$-local networks, non $n$-locality detection, etc have 
    undergone multifaceted analysis\cite{Hens,Kau,Fritz,Kau0,Chav,Ross,Kau1,Andr,Sau,And,Gisi,Car,Kau2,kundu,Tavakol,Kau3,Pod,Kau5,Poz,Kau6,Kau7,Kau8}. The study of quantum non-$n$-locality explores correlations beyond standard Bell nonlocality. Such network correlations reveal richer non-classicality of quantum systems thereby offering deeper insights into different network based quantum protocols\cite{Sun,Bau,Weh,lee,Koz}.

Like other quantum resources, nonlocality is fragile. Systems suffer from depletion of this resource, when subjected to environmental influence. Thus, it becomes extremely crucial to study channels which break nonlocality. Nonlocality breaking channels \cite{nlcbreak} were studied under the purview of the Bell-CHSH inequality \cite{bellchsh}. Precisely, the authors studied quantum channels in two qubit systems whose output was Bell-CHSH local \cite{nlcbreak}. As noted earlier, in the network scenario we witness nonlocality in myriad of variations, which bring notable distinction to studies on network nonlocality. In the present work, we study nonlocality breaking channels in the network scenario.

We have provided some criteria involving parameters of single-qubit channels. Non $n$-locality cannot be detected in quantum networks involving channels satisfying those criteria. Our probe is done for linear as well as star shaped n-local quantum networks. Interestingly, we have also pointed out presence of some resource(in terms of non $n$-locality) preserving single qubit  channels. We also study nonlocality breaking channels in the context of full network nonlocality. The entire study is done in the purview of some special inequalities in linear n-local networks as well as star networks. We have considered both unital and non-unital channels for this purpose and illustrative examples have been provided which further burtresses the work.
	
	Rest of our work is organised as follows. In sec.\ref{prelim}, we have provided necessary pre-requisites for our work. In sec.\ref{std0}, we have introduced the notion of non $n$-locality breaking channels. Characterization of non $n$-locality breaking channels in $n$-local networks is provided in sec.\ref{std1}(for linear topology) and then in sec.\ref{std2}(for star topology). Finally we also explore the role of  channels in breaking full network nonlocality in sec.\ref{fulln}. We have provided with some concluding remarks in sec.\ref{conc}.
	\section{Preliminaries}\label{prelim}
	
	\subsection{Bloch Matrix Representation}
	Let $\varrho$ be an arbitrary two qubit state shared between two distant parties Alice and Bob. In terms of Bloch parameters $\varrho $ can be written as follows\cite{Luo}:
	\begin{equation}\label{st4}
		\small{\varrho}=\small{\frac{1}{4} (\mathbb{I}_{2} \times \mathbb{I}_{2}  +\vec{m}.\vec{\sigma}\otimes \mathbb{I}_2+\mathbb{I}_2\otimes \vec{n}.\vec{\sigma}+\sum_{r_1,r_2=1}^{3}t_{r_1r_2}\sigma_{r_1}\otimes\sigma_{r_2})},
	\end{equation}
	with $\vec{\sigma}$$=$$ (\sigma_1,\sigma_2,\sigma_3), $ $\sigma_{j}$ representing Pauli operators along mutually perpendicular directions ($j$$=$$1,2,3$). $\vec{m}$$=$$ (m_1,m_2,m_3)$ and $\vec{n}$$=$$ (n_1,n_2,n_3)$ represent the local Bloch vectors ($\vec{m},\vec{n}$$\in$$\mathbb{R}^3$) of Alice and Bob respectively with $|\vec{m}|,|\vec{n}|$$\leq$$1.$ $ (t_{i,j})_{3\times3}$ denoting the correlation tensor $\mathcal{T}$ (a real matrix).\\
	The components of $\mathcal{T}$ are computed as follows: $t_{ij}$$=$$\textmd{Tr}[\varrho\,\sigma_{i}\otimes\sigma_{j}].$ \\
	$\mathcal{T}$ can be diagonalized by applying suitable local unitary operations \cite{Luo,Gam}:
	\begin{equation}\label{st41}
		\small{\varrho}^{'}=\small{\frac{1}{4} (\mathbb{I}_{2} \times \mathbb{I}_{2}+\vec{a}.\vec{\sigma}\otimes \mathbb{I}_2+\mathbb{I}_2\otimes \vec{b}.\vec{\sigma}+\sum_{j=1}^{3}w_{j}\sigma_{j}\otimes\sigma_{j})},
	\end{equation}
	Here the correlation tensor is $\textbf{W}$$=$$\textmd{diag} (w_{1},w_{2},w_{3}).$ $w_{1},w_{2},w_{3}$ are the eigenvalues of $\sqrt{\mathcal{T}^T\mathcal{T}},$ i.e., singular values of $\mathcal{T}.$
	
	\subsection{Quantum channels}
	Consider the Hilbert spaces ${\mathcal{H}}_A = {\mathcal{H}}_B = {\mathbb{C}}^d$. Let, $\mathbf{M}_d$ and $\mathbf{M}_k$ denote the set of $d \times d$ and  $k \times k$ complex matrices respectively. A linear map $\mathcal{N}: \mathbf{M}_d \rightarrow \mathbf{M}_d $ is said to be positive, if $\mathcal{N}(\rho) \ge 0$, for all $\rho \in \mathbf{M}_d$. A linear map $\mathcal{N}: \mathbf{M}_d \rightarrow \mathbf{M}_d$ is said to be $k$ positive if the induced map $\mathcal{N} \otimes id_B :\mathbf{M}_d \otimes \mathbf{M}_k \rightarrow \mathbf{M}_d \otimes \mathbf{M}_k$ is positive for some $k \in \mathbb{N}$. A linear map $\mathcal{N}: \mathbf{M}_d \rightarrow \mathbf{M}_d $ is completely positive if $\mathcal{N}: \mathbf{M}_d \rightarrow \mathbf{M}_d $ is positive for all $k \in \mathbb{N}$. To show that, a positive map $\mathcal{N}$ is completely positive, one makes use of the Choi-Jamiolkowski isomorphism \cite{jamiolkowski,choi}. The Choi matrix corresponding to the positive map $\mathcal{N}$ is $ {\mathcal{C}}_{\mathcal{N}} = (\mathcal{N} \otimes id_B) (\ket{{\phi}^+}\bra{{\phi}^+})$, where $\ket{{\phi}^+}=\frac{1}{\sqrt d} \sum_{i} \ket{ii}$ is the maximally entangled state in ${\mathbb{C}}^d\otimes {\mathbb{C}}^d$. Now, a positive map $\mathcal{N}$ is completely positive iff the Choi matrix ${\mathcal{C}}_{\mathcal{N}}$ corresponding to the map is positive semidefinite. Furthermore, a linear map $\mathcal{N}$ is said to be trace-preserving if $\text{Tr} (\mathcal{N} (\rho)) = \text{Tr}(\rho)$ for all $\rho \in  \mathbf{M}_d$. 
	
	As noted before, the evolution of quantum states are modeled through quantum channels, which are completely positive and trace-preserving maps (CPTP). The Kraus-Choi representation theorem \cite{kraus,choi,sudarshan} entails that any CPTP map $\mathcal{N}: \mathbf{M}_d \rightarrow \mathbf{M}_d$ can be written as,
	\begin{equation}
		\mathcal{N} (\rho) = \sum_{i} \mathcal{K}_{i} \hspace{0.05cm}\rho \hspace{0.05cm}{{\mathcal{K}}_{i}^{\dagger}}
	\end{equation}
	where, $\mathcal{K}_{i} \in \mathbf{M}_d $ are the Kraus operators corresponding to the channel $\mathcal{N}$ satisfying $\sum_{i} {{\mathcal{K}_{i}}^{\dagger}}  \mathcal{K}_{i} = I_d$. A channel is said to be \textit{unital} if it preserves the maximally mixed state i.e., $\mathcal{N} (I)=I$.
	
	For single qubit systems, channels have a special matrix representation. Since $\{ \mathbb{I}, \vec{\sigma} \}$ forms a basis for the matrices in $\mathbf{M}_2$, every single qubit channel can be represented by a $4 \times 4$ matrix 
	$ \mathbb{T}$, where $ \mathbb{T}$ is given by \cite{ruskaimap}, 
	\begin{equation}\label{nu1}
		\mathbb{T}=
		\begin{bmatrix}
			1 & \textbf{0} \\
			\vec{t} & T 
		\end{bmatrix}
	\end{equation}
	where is a $T$ is a $3 \times 3$ matrix (\textbf{0} and $\vec{t}$ and are row and column vectors, respectively). A single qubit channel is unital if and only if $\vec{t}=0$ \cite{ruskaimap}. A special subclass of non-unital single qubit channels is given by \cite{ruskaimap}, 
	\begin{equation}\label{nu2}
		\mathcal{N}_{NU}=
		\begin{bmatrix}
			1 & 0 & 0 & 0 \\
			0 & \lambda_1 & 0 & 0\\
			0 & 0 & \lambda_2 & 0\\
			t & 0 & 0 & \lambda_3
		\end{bmatrix}
	\end{equation}
	Parameters corresponding to $\mathcal{N}_{NU}$ must satisfy the following necessary conditions\cite{ruskaimap2}:
	\begin{eqnarray}\label{nu3}
		|\lambda_1|&\leq& 1\nonumber\\
		|\lambda_3|+|t|&\leq& 1\nonumber\\
		\lambda_1^2+t^2&\leq& (1-|\lambda_3|)^2\nonumber\\
		&&
	\end{eqnarray}
	
	An interesting class of channels are constituted by mixed unitary channels (also called random unitary channels), which can be presented as $\mathcal{N}_{mu} (\rho)= \sum_i p_i U_i \rho U^\dagger_i$, where $p_i$ is a probability distribution and $U_i$ are unitaries. While, mixed unitary channels are unital, the converse is not true in general. However, for single qubit systems they are equivalent. In fact, if $\mathcal{N}$ is a single qubit unital channel, then it can be written as the average of four unitary channels \cite{choiunital}: 
	\begin{eqnarray}\label{ruc}
		U_1& =& \begin{bmatrix}
			\alpha & \bar{\beta} \\
			-\beta & \bar{\alpha} 
		\end{bmatrix} \nonumber\\
		U_2& = &\begin{bmatrix}
			\alpha & -\bar{\beta} \\
			\beta & \bar{\alpha} 
		\end{bmatrix} \nonumber\\
		U_3& =& \begin{bmatrix}
			\bar{\alpha} & \beta \\
			-\bar{\beta} & \alpha 
		\end{bmatrix} \nonumber\\
		U_4& =& \begin{bmatrix}
			\bar{\alpha} & -\beta \\
			\bar{\beta} & \alpha 
		\end{bmatrix} .
	\end{eqnarray}
	Here, $\alpha, \beta \in \mathbb{C}$ with
	\begin{equation}\label{ch0}
		|\alpha|^2 + |\beta|^2 = 1	
	\end{equation}
	Entanglement breaking channels \cite{entbreak} are certain classes of quantum channels that completely disentangle the subsystem they are acting on from the remainder of the system. Precisely, a channel $\mathcal{N}_{EB}$ is entanglement breaking, if $(\mathcal{N}_{EB} \otimes id_B) (\rho)$ is separable for all $\rho$ \cite{entbreak}. A quantum channel  $\mathcal{N}_{EB}: \mathbf{M}_d \rightarrow \mathbf{M}_d $ is said to be entanglement breaking iff $(\mathcal{N}_{EB} \otimes id_B) (\ket{{\phi}^{+}} \bra{{\phi}^{+}})$ is separable, where, $\ket{{\phi}^{+}}=\frac{1}{\sqrt d} \sum_{i} \ket{ii}$ is the maximally entangled state in ${\mathbb{C}}^d\otimes {\mathbb{C}}^d$ \cite{entbreak}. It is known  \cite{entbreak} that an entanglement breaking channel can be written in terms of rank one Kraus operators. 
	
	Nonlocality breaking qubit channels were studied in \cite{nlcbreak}. A channel  $\mathcal{N}_{NB}$ is nonlocality breaking, if $(\mathcal{N}_{NB} \otimes id_B) (\rho)$ satisfies a certain Bell inequality for all input states $\rho$. The study in \cite{nlcbreak} was done under the ambit of the Bell-CHSH inequality. 
	
	\subsection{$n$-local Linear Network \cite{Kau}}\label{biloc}
		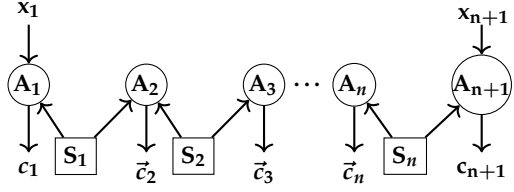
\begin{figure}
		\centering
		\begin{tikzpicture}[
			node distance=1.2cm and 1cm,
			every node/.style={font=\boldmath\bfseries\small},
			meas/.style={draw, circle, minimum size=0.55cm,inner sep=0pt,       
				align=center, 
				text height=1.2ex, 
				text depth=0ex},
			source/.style={draw, rectangle, minimum width=0.5cm, minimum height=0.3cm},
			arr/.style={->, thick},
			dots/.style={gray, thick}
			]
			
			\node[meas] (A1) at (0,0) {$\,\mathbf{A}_{1\,\,}$};
			\node[meas] (A2) [right=of A1] {$\,\mathbf{A}_{2\,\,}$};
			\node[meas] (A3) [right=of A2] {$\,\mathbf{A}_{3\,\,}$};
			\node at ($(A3)+(0.6,0)$) {\dots};
			\node[meas] (An) at ($(A3)+(1.2,0)$) {$\,\mathbf{A}_{n\,\,}$};
			\node[meas] (Anp1) [right=of An] {$\mathbf{\small{A_{n+1}}}$};
			
			\node[source] (S1) [below=0.7cm of $(A1)!0.4!(A2)$] {$\mathbf{S}_1$};
			\node[source] (S2) [below=0.7cm of $(A2)!0.4!(A3)$] {$\mathbf{S}_2$};
			\node[source] (Sn) [below=0.7cm of $(An)!0.4!(Anp1)$] {$\mathbf{S}_n$};
			
			\draw[arr] (S1) -- (A1);
			\draw[arr] (S1) -- (A2);
			
			\draw[arr] (S2) -- (A2);
			\draw[arr] (S2) -- (A3);
			
			\draw[arr] (Sn) -- (An);
			\draw[arr] (Sn) -- (Anp1);
			
			\draw[arr] (A1) -- ++(0,-0.9) node[below] {$c_1$};
			\draw[arr] (A2) -- ++(0,-0.9) node[below] {$\vec{c}_2$};
			\draw[arr] (A3) -- ++(0,-0.9) node[below] {$\vec{c}_3$};
			\draw[arr] (An) -- ++(0,-0.9) node[below] {$\vec{c}_n$};
			\draw[arr] (Anp1) -- ++(0,-0.9) node[below] {$\mathbf{\small{c_{n+1}}}$};
			
			\node at ($(A1)+(0,1)$) {$\mathbf{\small{x_{1}}}$};
			\draw[arr] ($(A1)+(0,0.8)$) -- (A1);
			
			\node at ($(Anp1)+(0,0.9)$) {$\,\mathbf{\small{x_{n+1}}}$};
			\draw[arr] ($(Anp1)+(0,0.8)$) -- (Anp1);\
		\end{tikzpicture}
		\caption{Schematic representation of linear $n$-local network}
		\label{nlocal1}
	\end{figure}
	In this network structure, $n$ independent sources are distributed among $n+1$ parties arranged in a linear pattern (see Fig.\ref{nlocal1}). $\forall i$$=$$1,2,...,n,$ let $\textbf{S}_i$ distribute a two particle state. Let hidden variable $\lambda_i$ characterize source $\textbf{S}_i.$ Sources being independent, distribution of $\lambda_1,\lambda_2,...,\lambda_n$ is factorizable:
	\begin{equation}\label{tr2}
		\rho (\lambda_1,...\lambda_n)=\Pi_{i=1}^n\rho_i (\lambda_i)
	\end{equation}
	$\rho (\lambda_1,...\lambda_n)$ denotes the probability density function of local hidden variables $\lambda_1,...,\lambda_n$ and $\rho_i (\lambda_i)$ is defined analogously.\\
	Eq.(\ref{tr2}) is referred to as \textit{$n$-local constraint}\cite{BRGP}.\\
	The measurement settings of the parties involved in the network are as follows:
	\begin{itemize}
		\item Each of the two extreme parties $\textbf{A}_1$ and $\textbf{A}_{n+1},$ receiving a single particle, perform any one of two projective measurements: $y_1$ and $y_{n+1}$ respectively ($y_1,y_{n+1}$$\in$$\{0,1$\}).
		\item $\forall i =2,3,...n,$ central party $\textbf{A}_{i}$ receives two qubits from $\textbf{S}_{i-1},\textbf{S}_{i}$ and performs a single measurement  having $4$ outcomes \cite{BRGP,Kau} on the state of the two qubits received. Let us denote corresponding linear network as $\textbf{N}_{n-lin}$(see Fig.\ref{nlocal1}).
	\end{itemize}
	$n+1$-partite correlations generated in $\textbf{N}_{n-lin}$ are $n$-local if those can be decomposed as\cite{Kau}:
	\begin{eqnarray}\label{tr1}
		P (b_1,\vec{b}_2,...,\vec{b}_{n},a_{n+1})=	\int...\int_{\Lambda_1,...\Lambda_n}d\lambda_1,...d\lambda_n \rho (\lambda_1,...\lambda_n) \textbf{P}_1,&&\nonumber\\
		\textbf{P}_1= P (b_1|y_1,\lambda_1) P (b_{n+1}|y_{n+1},\lambda_n)\Pi_{2,...,n}P (b_{i1},b_{i2}|\lambda_{i-1},\lambda_i)&&\nonumber\\
		\textmd{with } \vec{b}_i= (b_{i1},b_{i2}),\forall i \quad\quad&&
	\end{eqnarray}
	along with the $n$-local constraint given by Eq.(\ref{tr2}).
	$\vec{b}_i$$=$$ (b_{i1},b_{i2})$$\in$$\{ (0,0), (0,1), (1,0), (1,1)\}$ denote the $4$ outcomes of party $\textbf{A}_i (i$$=$$2,3,...,n).$ $b_1,b_{n+1}$$\in$$\{0,1\}$ denote binary outputs of $\textbf{A}_1$ and $\textbf{A}_{n+1}$ respectively.
	So, any set of $n+1$-partite correlations that do not satisfy both of these restrictions(Eqs.(\ref{tr1},\ref{tr2})) simultaneously, are termed as non $n$-local in $\textbf{N}_{n-lin}$.
	There exists $n$-local inequality($\textbf{I}_{n-lin}$,say) to detect non $n$-local correlations \cite{Kau}:
	\begin{eqnarray}\label{ineqb}
		&& \sqrt{|I_n|}+\sqrt{|J_n|}\leq  1,\,  \textmd{where}\\
		&& I_n=\frac{1}{4}\sum_{y_1,y_{n+1}}\langle D_{1,y_1}D_2^0D_3^0...D_{n}^0D_{n+1,y_{n+1}}\rangle\nonumber\\
		&&   J_n= \frac{1}{4}\sum_{y_1,y_{n+1}} (-1)^{y_1+y_{n+1}}\langle \small{ D_{1,y_1}D_2^1...D_{n}^1D_{n+1,y_{n+1}}}\rangle\,\,\textmd{with} \nonumber\\
		&&   \langle D_{1,y_1}D_2^i...D_{n}^iD_{n+1,y_{n+1}}\rangle = \sum_{\textbf{D}_1} (-1)^{b_1+b_{n+1}+\sum_{j=2}^{n}b_{j (i+1)}}Q_1,\nonumber\\
		&& \textmd{\small{where}}\,Q_1=\small{p (b_1,
			\vec{b}_2,...,\vec{b}_{n},
			b_{n+1}|y_1,y_{n+1})},\, i=0,1\nonumber\\
		&& \textbf{D}_1=\{b_1,b_{21},b_{22},...,
		b_{n1},b_{n2},b_{n+1}\}\nonumber
	\end{eqnarray}
	Violation of $\textbf{I}_{n-lin}$ (Eq.(\ref{ineqb})) guarantees non $n$-locality of corresponding measurement correlations. However, there is no definite conclusion in case correlations satisfy $\textbf{I}_{n-lin}.$\\
	Now, $\forall i,$ let $\textbf{S}_i$ distribute any two qubit state $\rho_{i}$(Eq.\ref{st41}).
	Let each of $\textbf{A}_2,...,\textbf{A}_n$ perform Bell basis measurement(BSM) and $\textbf{A}_1$ and $\textbf{A}_{n+1}$ each perform qubit projective measurement in any one of two directions. \\
	For above measurements, the upper bound of Eq.(\ref{ineqb}) is given by\cite{Gisi}:
	\begin{equation}\label{boundlin}
		\textbf{B}_{n-lin}=\sqrt{\Pi_{j=1}^n E_{j1}+\Pi_{j=1}^n E_{j2}}
	\end{equation}
	with $E_{i1},E_{i2},E_{i3}$$\in$$\{w_{i1},w_{i2},w_{i3}\}$ such that $E_{i3}$$\leq$$E_{i2}$$\leq$$E_{i1}$ denote the ordered eigen values of $\sqrt{ (W_i)^TW_i}, \forall i.$\\
	
	$\textbf{I}_{n-lin}$ is violated if:
	\begin{equation}\label{up11}
		\textbf{B}_{n-lin}>1
	\end{equation}
	
	\subsection{$n$-local Star Network\cite{Tava}}\label{star1}
		\begin{figure}
		\begin{tikzpicture}[
			node distance=1.4cm and 1.6cm,
			every node/.style={font=\boldmath\bfseries\small},
			meas/.style={
				draw, circle, minimum size=0.6cm,
				inner sep=0pt, align=center,
				text height=1.2ex, text depth=0ex
			},
			source/.style={
				draw, rectangle, minimum width=0.5cm, minimum height=0.3cm,
				inner sep=1pt, align=center
			},
			arr/.style={->, thick}
			]
			
			\node[meas] (A1) at (0,0) {$\mathbf{A}_1$};
			
			\node[meas] (A2) at (150:3cm) {$\mathbf{A}_2$};
			\node[meas] (A3) at (120:3cm) {$\mathbf{A}_3$};
			
			\node[meas] (An) at (60:3cm) {$\mathbf{A}_n$};
			\node[meas] (Anp1) at (30:3cm) {$\mathbf{A}_{n+1}$};
			
			\node at (90:3.2cm) {\dots};
			
			\node[source] (S1) at ($(A1)!0.5!(A2)$) {$\mathbf{S}_1$};
			\node[source] (S2) at ($(A1)!0.5!(A3)$) {$\mathbf{S}_2$};
			\node[source] (Sn) at ($(A1)!0.5!(An)$) {$\mathbf{S}_{n-1}$};
			\node[source] (Snp1) at ($(A1)!0.5!(Anp1)$) {$\mathbf{S}_n$};
			
			\foreach \X in {S1, S2, Sn, Snp1} {
				\draw[arr] (\X) -- (A1);
			}
			\draw[arr] (S1) -- (A2);
			\draw[arr] (S2) -- (A3);
			\draw[arr] (Sn) -- (An);
			\draw[arr] (Snp1) -- (Anp1);
			
			\foreach \i/\A in {2/A2, 3/A3, n/An, {n+1}/Anp1} {
				\node at ($(\A)+(0,0.9)$) {$x_{\i}$};
				\draw[arr] ($(\A)+(0,0.7)$) -- (\A);
				\draw[arr] (\A) -- ++(0,-0.8) node[below] {$c_{\i}$};
			}
			
			\draw[arr] (A1) -- ++(0,-0.6) node[below] {$\bar{c}_1$};
		\end{tikzpicture}
		\caption{Schematically representing $n$-local network in star topology}
		\label{nlocal2}
	\end{figure}
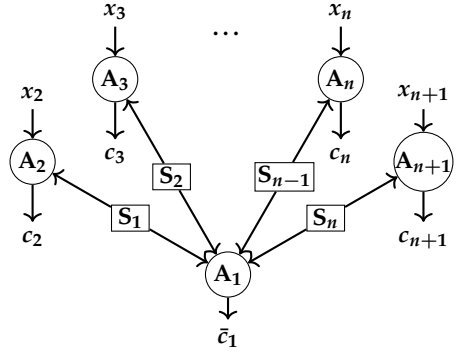
	This network is a non-linear $n$-local network that consists of one central party $\textbf{A}_1$ and $n$ number of edge(extreme) parties $\textbf{A}_2,...,\textbf{A}_{n+1}$ (see Fig.\ref{nlocal2}). Let the network be denoted by $\textbf{N}_{n-star}.$\\
	Each of the edge parties receives one particle from a source. Each of the sources $\textbf{S}_i$ is characterized by variable $\lambda_i(i$$=$$1,2,...,n)$ The sources being independent, joint distribution of the variables $\lambda_1,...,\lambda_n$ is factorizable:
	\begin{equation}\label{tr1}
		q(\lambda_1,\lambda_2...\lambda_n)=\Pi_{i=1}^n q_i(\lambda_i),
	\end{equation}
	with $q_i$ denotes the normalized distribution of $\lambda_i,\forall i.$ Source independence condition(Eq.(\ref{tr1})) represents the $n$-local constraint\cite{Tava}. \\
	Each of the edge parties $\textbf{P}_i$ chooses to perform from any one of two possible measurements($x_i$$\in$$\{0,1\}$) whereas the central party performs a fixed measurement. $n+1$-partite measurement correlations are local if:
	\begin{eqnarray}\label{tr2s}
		&&\small{p(\vec{\mathfrak{c}}_1,c_2,c_3...,c_{n+1}|x_2,x_3,...,,x_{n+1})}=\nonumber\\
		&&\int_{\Lambda_1}\int_{\Lambda_2}...\int_{\Lambda_n}
		d\lambda_1d\lambda_2...d\lambda_n\,q(\lambda_1,\lambda_2,...\lambda_n) \textbf{P} ,\,\textmd{with}\nonumber\\
		&&\textbf{P}=p(\vec{\mathfrak{c}}_1|\lambda_1,...,\lambda_n)\Pi_{i=1}^{n} p(c_{i+1}|x_{i+1},\lambda_i)\nonumber\\
		&&
	\end{eqnarray}
	$n+1$-partite correlations satisfying both Eqs.(\ref{tr1},\ref{tr2}) are termed as $n$-local correlations. Hence, any set of correlations that do not satisfy Eqs.(\ref{tr1},\ref{tr2}) simultaneously, are termed as non $n$-local\cite{BGP}.\\ 
	The existing $n$-local inequality for $\textbf{N}_{n-star}$ is given by\cite{Kau}:
	\begin{eqnarray}\label{ineqs}
		&&\frac{1}{2^{n-2}} \sum_{i=1}^{2^{n-1}}|J_{i}|^{\frac{1}{n}}\leq  1,\,  \textmd{where}\\
		&& J_i=\frac{1}{2^n}\sum_{x_2,...,x_{n+1}}(-1)^{g_i(x_2,...,x_{n+1})}\langle \textbf{A}_{(1)}^{(i)}\textbf{A}_{x_2}^{(2)}...\textbf{A}_{x_{n+1}}^{(n+1)}\rangle\nonumber\\
		&&\langle \textbf{A}_{(1)}^{(i)}\textbf{A}_{x_2}^{(2)}...\textbf{A}_{x_{n+1}}^{(n+1)}\rangle=\sum_{\textbf{D}_1}(-1)^{\tilde{\mathfrak{c}}_1^{(i)}+\textbf{c}_2+...
			+\textbf{c}_{n+1}}N_1,\nonumber\\
		&& \textmd{\small{where}}\,N_1=\small{p(\tilde{\textbf{c}}_1,
			\textbf{c}_2,...,\textbf{c}_{n+1}
			|x_2,...,x_{n+1})}\,\textmd{\small{and}}\nonumber\\
		&&\textbf{D}_1=\{\textbf{c}_{11},....,\textbf{c}_{1n},\textbf{c}_2,....
		,\textbf{c}_{n+1}\}\nonumber\\
	\end{eqnarray}
	In Eq.(\ref{ineqs}), $\forall i$$=1,2,...,2^{n-1},$ $\tilde{\mathfrak{c}}_1^{(i)}$ represents an output bit generated by classical post-processing of the raw output string $\overline{\textbf{c}}_1$$=$$(\textbf{c}_{11},....,\textbf{c}_{1n})$ of $\textbf{P}_1.$ In Eq.(\ref{ineqs}), $\forall i,\, g_i$ are functions of the input variables $x_2,...,x_{n+1}$ of the extreme parties\cite{Tava}. Each $g_i$ contains an even number of $x_2,...,x_{n+1}.$ \\
	Let each of the sources distribute an arbitrary two qubit state (Eq.\ref{st41}). Under above measurement settings, the upper bound of Eq.(\ref{ineqs}) is given by $	B_{n-star}$\cite{kundu} where: 
	\begin{equation}\label{boundstar}
		B_{n-star}=	\sqrt{(\Pi_{i=1}^n  E_{i1})^{\frac{2}{n}}+(\Pi_{i=1}^nE_{i2})^{\frac{2}{n}}}.
	\end{equation}
	$n$-local inequality (Eq.(\ref{ineqs})) is thus violated if:
	\begin{equation}\label{up211}
		B_{n-star}>1.
	\end{equation}
	
	\subsection{Full Network Nonlocality}
	The concept of full network nonlocality was introduced in \cite{Poz}. In any network scenario, measurement correlations are termed \textit{fully network nonlocal} if and only if  the correlations cannot be modeled by any $n$-local hidden variable(HV) model such that at least one of $n$ independent sources in the network is of local-variable nature whereas remaining $n$$-$$1$ sources can be independent nonlocal resources.\\
	In general we can consider arbitrary $n$-local network. For our purpose, we consider  $n$-local star network. To be precise, central party $A_1$ performs single measurement whereas each of $\textbf{A}_2,..,\textbf{A}_{n+1}$(edge parties) perform from a set of two arbitrary dichotomic projective measurements. \\
	Corresponding measurement correlation $p(\bar{a}_1,a_2,...,a_{n+1}|x_2,...,x_{n+1})$ is \textit{not full network nonlocal} if it can be factorized as:
	\begin{eqnarray}\label{fnn1}
		P(\vec{\mathfrak{c}}_1,c_2,c_3...,c_{n+1}|x_2,...,x_{n+1})=\sum_{\lambda}\Lambda(\lambda)
		P(c_j|x_j,\lambda)R\,\,\textmd{where}\nonumber\\
		R=P(\vec{\mathfrak{c}}_1,c_2,...c_{j-1},c_{j+1},...,c_{n+1}|x_2,..,x_{k-1},x_{k+1},...,x_{n+1},\lambda)\nonumber	\\
	\end{eqnarray}
	$\Lambda(\lambda)$ represents the probability distribution of the local hidden variable $\lambda$ that characterizes $j^{th}$ source $\textbf{S}_j$ shared in between $\textbf{A}_1$ and $\textbf{A}_j.$\\
	Eq.(\ref{fnn1}) implies that for any $j$$\in$$\{1,2,,...,n$\}, $j^{th}$ source is modeled by a local hidden variable $\lambda.$ So it is clear that if at least one of $n$ sources can be modeled by a local hidden variable, then even if rest of $n$$-$$1$ sources are maximally nonlocal(no-signalling box), corresponding network correlations fail to be fully network nonlocal.\\
	\subsubsection{Detection of full network nonlocality for $n$$=$$3$}
	In trilocal star network, Eq.(\ref{ineqs}) with an escalated trilocal bound $\sqrt[3]{2}$(instead of $1$) was established as a trilocal inequality for detecting full network non trilocality\cite{Poz}. Violation of Eq.(\ref{ineqs}) with bound $\sqrt[3]{2}$ thus guarantees generation of full network non trilocality. To be precise, violation of the following trilocal inequality implies corresponding correlations to be fully network nonlocal:
	\begin{eqnarray}\label{ineqsn}
		&&\frac{1}{2} \sum_{i=1}^{4}|J_{i}|^{\frac{1}{n}}\leq  \sqrt[3]{2},\,  \textmd{where}\\
		&& J_i=\frac{1}{2^3}\sum_{x_2,...,x_{4}}(-1)^{g_i(x_2,...,x_{4})}\langle \textbf{A}_{(1)}^{(i)}\textbf{A}_{x_2}^{(2)}...\textbf{A}_{x_{4}}^{(4)}\rangle\nonumber\\
		&&\langle \textbf{A}_{(1)}^{(i)}\textbf{A}_{x_2}^{(2)}...\textbf{A}_{x_{4}}^{(4)}\rangle=\sum_{\textbf{D}_1}(-1)^{\tilde{\mathfrak{c}}_1^{(i)}+\textbf{c}_2+...
			+\textbf{c}_{4}}N_1,\nonumber\\
		&& \textmd{\small{where}}\,N_1=\small{p(\tilde{\textbf{c}}_1,
			\textbf{c}_2,...,\textbf{c}_{4}
			|x_2,...,x_{4})}\,\textmd{\small{and}}\nonumber\\
		&&\textbf{D}_1=\{\textbf{c}_{11},\textbf{c}_{13},\textbf{c}_2,\textbf{c}_{4}\}\nonumber\\
	\end{eqnarray}
	Terms used above are already explained after Eq.(\ref{ineqs}). As the correlators involved in Eq.(\ref{ineqs} for $n$$=$$3$) are same as that in Eq.(\ref{ineqsn}), following criterion is sufficient to detect full network nonlcality in trilocal star network:
	\begin{eqnarray}\label{up212}
		\sqrt{(\Pi_{i=1}^3  E_{i1})^{\frac{2}{3}}+(\Pi_{i=1}^3E_{i2})^{\frac{2}{3}}}>\sqrt[3]{2}
	\end{eqnarray}
	 
	\subsection{Standard and Non Standard Networks}
	For ease of understanding we may consider any $n$-local network as \textit{standard $n$-local network} if any of $n$ independent sources can be characterized by local hidden variable model. On the other hand if none of the sources can be of local variable nature\cite{Poz}, then corresponding network will be referred to as \textit{non standard n-local network}. It is clear that full network nonlocality can be exploited only in non standard $n$-local network. As most of our results will be discussed in standard $n$-local network, so we will call any standard $n$-local network as $n$-local network only. Otherwise, we will use the term non-standard $n$-local network for e.g. in sec.\ref{fulln} where we will provide observations encompassing full network nonlocality.
	\section{Non $n$-locality Breaking Channels}\label{std0}
	To provide our observations in the context of loss in non $n$-local resources due to communication through channels, we define the notion of non $n$-locality breaking channels in this section. \\
	\subsection{Non $n$-local Resource}
	Any network structure involves more than one quantum source. Hence, unlike standard Bell nonlocality, for generation of non $n$-local correlations, one needs more than one quantum state. Now, for the specific network scenarios considered here, each source distributes a two-qubit state between two distant nodes(parties). \\
	Consider the case where each of the $n$ independent sources $\textbf{S}_1,...,\textbf{S}_n$ sends two-qubit states $\varrho_1,...,\varrho_n$ respectively.
	Let $\varrho_{in}$ denote the overall state in the network:
	\begin{equation}\label{tr4}
		\varrho_{in}=\otimes_{j=1}^n \varrho_{j}.
	\end{equation}
	With respect to present scenario we formally define \textit{non $n$-local resource} as follows:
    \begin{definition}
	$\varrho_{in}$ is said to be non $n$-local resource if measurement correlations generated in the network involving $\varrho_{in}$ are non $n$-local.
    \end{definition}
	Violation of any $n$-local inequality, compatible with the network topology, acts as a sufficient criterion to  detect non $n$-local resource(if any) in the network. \\
	
	\subsection{Non $n$-locality Breaking Channel}
	As each of $n$ sources is distributing a two-qubit state, total $2n$ qubits are distributed across the entire network. Let $\varrho_i^{(1)},\varrho_i^{(2)}$ denote two qubits constituting the two-qubit state $\varrho_i$($i$$=$$1,2,...,n$). Distribution of the qubits among the nodes depends on the specific network topology(see Figs.\ref{nlocalc1},\ref{nlocalc2}). \\
	For ease of discussion, let us use a numbering of $2n$ qubits as follows:
	\begin{eqnarray}\label{nb1}
		\varrho_i^{(1)}&\rightarrow& Q_{2i-1}\nonumber\\
		\varrho_i^{(2)}&\rightarrow& Q_{2i},\,\,\forall i=1,2,...,n
	\end{eqnarray}
	Each qubit is thus individually communicated from a source to a node. In course of communication, it may come under the effect of environmental noise. For that we may consider that a channel is operating over the qubit. So to demonstrate destruction of non $n$-locality, one need to consider only the effect(if any) of single qubit quantum channels over some or all the qubits distributed across the nodes in the network. For our purpose, we put forward the idea of $k$$-$$use$ of a single qubit channel in a network.\\
	\begin{definition}
	$\textbf{k}$-use of a channel: A single-qubit channel is said to be used $k$ times in a network if $k$ number of qubits are communicated through that channel in the network.
    \end{definition}
	Let $\mathcal{N}$ denote a single-qubit channel. Let $\varrho_{in}^{'}$ denote the overall state across the network after action of $k$ use of $\mathcal{N}.$ When $\mathcal{N}$ is used $k$ times then anyone of the following cases may occur:
	\begin{enumerate}\label{enu1}
		\item[i] Both qubits of some or all $\rho_i$ get communicated individually through $\mathcal{N}.$\\
		W.L.O.G., let $\mathcal{N}$ act on each of the qubits of each of $\rho_1,\rho_2,...,\rho_m(m$$\leq$$n)$ So here $k$$=$$2m.$ In this case $\varrho_{in}^{'}$ is given by:
		\begin{eqnarray}\label{nb2}
			\varrho_{in}^{'}&=&\otimes_{j=1}^m(\mathcal{N}\otimes \mathcal{N}(\varrho_j))\otimes_{j=m+1}^n\varrho_j
		\end{eqnarray}
		\item[ii] Single qubit of some or all $\rho_i$ get communicated individually through $\mathcal{N}.$\\
		W.L.O.G., let $\mathcal{N}$ acts on one of the two qubits of each of $\rho_1,\rho_2,...,\rho_m(m$$\leq$$n)$ So here $k$$=$$m.$ In this case $\varrho_{in}^{'}$ is given by:
		\begin{eqnarray}\label{nb3}
			\varrho_{in}^{'}&=&\otimes_{j=1}^m(\mathcal{N}\otimes id(\varrho_j))\otimes_{j=m+1}^n\varrho_j\\
			&&id \textmd{ \small{denote identity channel}}\nonumber
		\end{eqnarray}
		\item[iii] Single qubit of some states and both qubits of some other states get communicated individually through $\mathcal{N}.$\\
		W.L.O.G., let $\mathcal{N}$ acts on one of the two qubits of each of $\rho_1,\rho_2,...,\rho_{m_1}(m_1$$\leq$$n)$ and individually on both qubits of $\rho_{m_1+1},\rho_{m_1+2},...,\rho_{m_1+m_2}(m_1$$+$$m_2$$\leq$$n)$ So here:
		\begin{equation}\label{k-val}
			k=m_1+2m_2.
		\end{equation}
		In this case $\varrho_{in}^{'}$ is given by:
		\begin{eqnarray}\label{nb4}
			\varrho_{in}^{'}&=&\otimes_{j=1}^{m_1}(\mathcal{N}\otimes  id (\varrho_j))\otimes_{j=m_1+1}^{m_1+m_2}(\mathcal{N}\nonumber\\
			&&\otimes \mathcal{N}(\varrho_j))\otimes_{j=m_1+m_2+1}^n\varrho_j
		\end{eqnarray}
	\end{enumerate}
	It can be seen that with $(m_1,m_2)$$\neq$$(0,0),$ third possibility encompasses all the possibilities:
	\begin{itemize}
		\item If $m_1$$=$$0,$ hence $k$$=$$2m_2,$ then we get case.(i).
		\item	If $m_2$$=$$0,$ hence $k$$=$$m_1,$  then we get case.(ii).
		\item For $m_1,m_2$ both non zero, we get case.(iii).\\
	\end{itemize}
	The notion of $k$-use of channel $\mathcal{N}$ will aid in illustrating the effect of $\mathcal{N}$ in destroying the resource of non $n$-locality in a quantum network. We next put forward a formal definition of \textit{$k$-use non $n$-locality breaking channel}.
    \begin{definition}
	A single qubit channel $\mathcal{N}$ is said to be $k$-use non $n$-locality breaking channel if for any $n$ independent sources distributing arbitrary two-qubit states in a $n$-local network, overall state resulting from communication of $k$ qubits through $\mathcal{N}$ fails to be a non $n$-local resource.
    \end{definition}
	Consider any $n$-local network with $\varrho_{in}$  denoting arbitrary overall state in the network. So $\varrho_{in}$ characterize arbitrary quantum sources involved in the network, i.e., arbitrary two-qubit states $\varrho_1,...,\varrho_n$ generated from $\textbf{S}_1,...,\textbf{S}_n$ before being communicated to the nodes. Let $\mathcal{N}$ be used $k$ times in the network. Let $\varrho_{in}^{'}$ denote the resulting overall state. So $\varrho_{in}^{'}$ characterize evolution(if any) of $\varrho_1,...,\varrho_n$ when $k$ out of $2n$ qubits are acted upon by $\mathcal{N}$ individually. $\mathcal{N}$ will be $k$-use non $n$-locality breaking channel if $\varrho_{in}^{'}$ fails to generate non $n$-local correlations. For more preciseness, let us refer $\varrho_{in}$(Eq.(\ref{tr4})) as \textit{input state of the network} and $\varrho_{in}^{'}$(given by anyone of Eqs.(\ref{nb2},\ref{nb3},\ref{nb4})) as \textit{output state of the network} indicating that $\varrho_{in}^{'}$ results due to distribution of $k$ qubits through channel $\mathcal{N}$ in the network. $\mathcal{N}$ will be $k$-use non $n$-locality breaking channel if for arbitrary input state $\varrho_{in},$ the output state $\varrho_{in}^{'}$ is $n$-local.
	\subsection{Characterizing $k$-Use Non $n$-Locality Breaking Channel}
	Let us consider any $n$-local network involving arbitrary quantum sources, i.e., considering $n$ number of arbitrary two-qubit states in the network. Let there be $k$-use of $\mathcal{N}.$ If resulting $n+1$-partite correlations admit $n$-local HV model then $\mathcal{N}$ is $k$-use non $n$-locality breaking channel. However, for obvious complexity in constructing $n$-local model for any network correlations, we adhere to some practically feasible strategy for illustrating role of a channel as a non $n$-locality breaking channel. \\
	Let $\mathbf{I}$ denote a $n$-local inequality compatible with the network's topology. Let there be $k$-use of $\mathcal{N}.$ If resulting $n+1$-partite correlations fail to violate $\mathbf{I},$ then that ensures $\mathcal{N}$ as $k$-use non $n$-locality breaking channel with respect to $\mathbf{I}.$ For rest of our discussion, we will refer any  $k$-use non $n$-locality breaking channel with respect to some $n$-local inequality $\mathbf{I},$ as \textit{$k$-use non $n$-locality breaking channel} only.\\
	With the above strategy, we next proceed to characterize $k$-use non $n$-locality breaking channel in some well-known network topologies. Precisely, we will consider linear $n$-local network and $n$-local star network.
	\section{Characterization in Linear $n$-local Network($\textbf{N}_{lin}$)}\label{std1}
		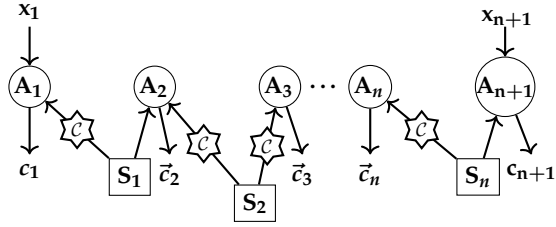
\begin{figure}
		\centering
		\begin{tikzpicture}[
			node distance=1.3cm and 1.1cm,
			every node/.style={font=\boldmath\bfseries\small},
			meas/.style={draw, circle, minimum size=0.55cm,inner sep=0pt,       
				align=center, 
				text height=1.2ex, 
				text depth=0ex},
			source/.style={draw, rectangle, minimum width=0.5cm, minimum height=0.3cm},
			arr/.style={->, thick},
			chlabel/.style={
				draw, star, star points=6, fill=white,
				inner sep=0.8pt, font=\scriptsize\bfseries
			},
			dots/.style={gray, thick}
			]
			
			\node[meas] (A1) at (0,0) {$\,\mathbf{A}_{1\,\,}$};
			\node[meas] (A2) [right=of A1] {$\,\mathbf{A}_{2\,\,}$};
			\node[meas] (A3) [right=of A2] {$\,\mathbf{A}_{3\,\,}$};
			\node at ($(A3)+(0.6,0)$) {\dots};
			\node[meas] (An) at ($(A3)+(1.2,0)$) {$\,\mathbf{A}_{n\,\,}$};
			\node[meas] (Anp1) [right=of An] {$\mathbf{\small{A_{n+1}}}$};
			
			\node[source] (S1) [below=0.95cm of $(A1)!0.8!(A2)$] {$\mathbf{S}_1$};
			\node[source] (S2) [below=1.3cm of $(A2)!0.8!(A3)$] {$\mathbf{S}_2$};
			\node[source] (Sn) [below=0.95cm of $(An)!0.8!(Anp1)$] {$\mathbf{S}_n$};
			
			\draw[arr] (S1) --node[midway,chlabel] {$\mathcal{C}$} (A1);
			\draw[arr] (S1) -- (A2);
			
			\draw[arr] (S2) --node[midway,chlabel] {$\mathcal{C}$} (A2);
			\draw[arr] (S2) --node[midway,chlabel] {$\mathcal{C}$} (A3);
			
			\draw[arr] (Sn) --node[midway,chlabel] {$\mathcal{C}$} (An);
			\draw[arr] (Sn) -- (Anp1);
			
			\draw[arr] (A1) -- ++(0,-0.9) node[below] {$c_1$};
			\draw[arr] (A2) -- ++(0.2,-0.9) node[below] {$\vec{c}_2$};
			\draw[arr] (A3) -- ++(0.3,-0.9) node[below] {$\vec{c}_3$};
			\draw[arr] (An) -- ++(0,-0.9) node[below] {$\vec{c}_n$};
			\draw[arr] (Anp1) -- ++(0.35,-0.9) node[below] {$\mathbf{\small{c_{n+1}}}$};
			
			\node at ($(A1)+(0,1)$) {$\mathbf{\small{x_{1}}}$};
			\draw[arr] ($(A1)+(0,0.8)$) -- (A1);
			
			\node at ($(Anp1)+(0,0.9)$) {$\,\mathbf{\small{x_{n+1}}}$};
			\draw[arr] ($(Anp1)+(0,0.8)$) -- (Anp1);\
		\end{tikzpicture}
		\caption{Single-qubit channel $\mathcal{C}$ acting on some of the qubits in a linear $n$-local network $\textbf{N}_{lin}.$ As shown here only one qubit from each of the sources $\mathbf{S}_1$ and $\mathbf{S}_n$ and both qubits from $\mathbf{S}_2$ are passing through $\mathcal{C}.$  }
		\label{nlocalc1}
	\end{figure}
	
	We have already discussed about linear network scenario in details(see sec.\ref{prelim}). Let $\varrho_{in}$(Eq.(\ref{tr4})) denote the  input state to $\textbf{N}_{lin}.$
	Let $\mathcal{N}$ denote an arbitrary single qubit channel. Let $\mathcal{N}$ be used $k$ times in $\textbf{N}_{lin}$(see Fig.\ref{nlocalc1} for example) So $k$$\leq$$2n.$ \\
	Let $\Theta_{\mathcal{N}}$ denote the set of parameters describing $\mathcal{N}.$ If a qubit is passed through $\mathcal{N},$ then the state parameters corresponding to the resulting qubit involve $\Theta_{\mathcal{N}}.$ In $\textbf{N}_{lin}$, characterizing $\mathcal{N}$ as a $k$-use non $n$-locality breaking channel thus results in deriving constraints over $\Theta_{\mathcal{N}}$ for which $\mathcal{N}$ destroys non $n$-locality of arbitrary input state $\varrho_{in}$, i.e., $\varrho_{in}^{'}$ satisfies $\mathbf{I}_{lin}$(Eq.(\ref{ineqb})). 
	\subsection{Unital Channel}
	Let $\mathcal{N}_U$ be  a single-qubit unital channel. As already pointed out in sec.\ref{prelim}, dimension being $2,$ $\mathcal{N}_U$ can be invariably considered as a random unitary channel with $p_i$$=$$\frac{1}{4},i$$=$$1,2,3,4$. 
    
	Being a single-qubit random unitary channel, $\Theta_{\mathcal{N}_{U}}$ consists of two parameters $\Theta_{\mathcal{N}_{U}}$$=$$\{\alpha,\beta\}.$ Let any such $\mathcal{N}_{U}$ be used $k$ times in $\textbf{N}_{lin}.$ In this context an obvious query rises: \textit{when will $\mathcal{N}_U$ turn out to be a non $n$-locality breaking channel?} We provide response to this query by deriving restrictions over the channel parameters $\alpha,\beta.$ The theorems below formalize our observations.\\
	Before we put forward our observation, let us set the following convention:\\
	\textit{Any complex number  $C$$=$$c_1$$+$$\imath$$c_2$ is said to be proper if it is neither real nor purely imaginary, i.e.: }
	\begin{eqnarray}\label{complex}
		Re(C)=c_1&\neq& 0\,\,\textmd{and} \nonumber\\
		Im(C)=c_2&\neq& 0; \nonumber
	\end{eqnarray}
	\textit{In case any one of the above criteria in Eq.(\ref{complex}) is violated, then we call corresponding complex number $C$ as improper complex number.}
	\begin{theorem}\label{thec1}
		If any single-qubit unital channel $\mathcal{N}_{U}$ be such that both channel parameters $\alpha,\beta$ are proper complex numbers satisfying
		\begin{eqnarray}\label{resu1}
			\textmd{Max}[(Re(\alpha))^2,(Re(\beta))^2,(Im(\alpha))^2,(Im(\beta))^2]-\nonumber\\
			\textmd{Min}[(Re(\alpha))^2,(Re(\beta))^2,(Im(\alpha))^2,(Im(\beta))^2]\leq \frac{1}{2^{1+\frac{1}{k}}},\nonumber\\
		\end{eqnarray}
		then $\mathcal{N}_{U}$ is $k$-use non $n$-locality breaking channel with respect to $n$-local inequality Eq.(\ref{ineqb}) in any linear $n$-local network.\\
		In particular, if $\mathcal{N}_{U}$ be such that any three of  $Re(\alpha),Im(\alpha),Re(\beta),Im(\beta)$ are identical for proper complex numbers $\alpha,\beta$ and also these parameters satisfy:
		\begin{eqnarray}\label{resu1o}
			\textmd{Max}[(Re(\alpha))^2,(Re(\beta))^2,(Im(\alpha))^2,(Im(\beta))^2]-\nonumber\\
			\textmd{Min}[(Re(\alpha))^2,(Re(\beta))^2,(Im(\alpha))^2,(Im(\beta))^2]\leq \frac{1}{2^{\frac{1}{k}}},\nonumber\\
		\end{eqnarray} 
		then such $\mathcal{N}_{U}$ is $k$-use non $n$-locality breaking channel with respect to $n$-local inequality Eq.(\ref{ineqb}) 
	\end{theorem}
	\textbf{Proof:} See Appendix.\ref{appenc1}.\\
	On restricting ourselves to detection of non $n$-locality via violation of Bell-type inequality, Theorem.\ref{thec1} gives a sufficient criterion for a single-qubit random unitary, hence a unital channel to be non $n$-locality breaking. However the criterion is not necessary as we cannot arrive at a definite conclusion in case $\mathcal{N}_{U}$ does not satisfy any of Eqs.(\ref{resu1},\ref{resu1o}). As is expected, the criterion depends on the count($k$) of qubits coming under action of the channel in $\mathbf{N}_{lin}.$\\ 
	In above theorem we have prescribed a set of criteria over channel parameters which when satisfied guarantees corresponding unital channel(used $k$ times) to be destroying non $n$-local resource. To this end it may be noted that in the above set of criteria we have considered both $\alpha,\beta$ to be proper complex numbers. It turns out that if we relax this restriction then corresponding channel no longer destroys non $n$-local resource in $\mathbf{N}_{lin}$. The theorem below justifies our claim.
	\begin{theorem}\label{thec2}
		If any single-qubit unital channel $\mathcal{N}_{U}$ be such that both the channel parameters $\alpha,\beta$ are improper complex numbers or any one of them is null, then $\forall k$$=$$1,2,...,2n,$  $\mathcal{N}_{U}$ is not a $k$-use non $n$-locality breaking channel in any linear $n$-local network.
	\end{theorem}
	\textbf{Proof:} See Appendix.\ref{appenc2}.\\
	Let any $\mathcal{N}_{U},$ as prescribed in Theorem.\ref{thec2}, be used $k$ times in $\mathbf{N}_{lin}.$ Let $n$ arbitrary  two-qubit states $\varrho_1,..,\varrho_n$ be distributed in $\mathbf{N}_{lin}$ such that $k$($\leq$$2n$) qubits are passed through $\mathcal{N}_{U}.$ Then for at least one set of input states $\varrho_1,..,\varrho_n,$ resulting overall state $\varrho_{in}^{'}$ will generate non $n$-local correlations in $\mathbf{N}_{lin}.$
	Any $\mathcal{N}_{U}$ satisfying theorem.\ref{thec2} thus acts as a non $n$-locality-preserving channel. From practical point viewpoint any such channel is important as it ensures that the intrinsic quantum correlations(non $n$-locality) of entangled states are not destroyed during transmission or interaction. \\
	Let us now use above theorems to illustrate the role of some well-known unital single-qubit channels.
	\subsubsection{\textbf{Depolarizing Channel}}
\begin{center}
		\begin{figure}
			\includegraphics[width=3.4in]{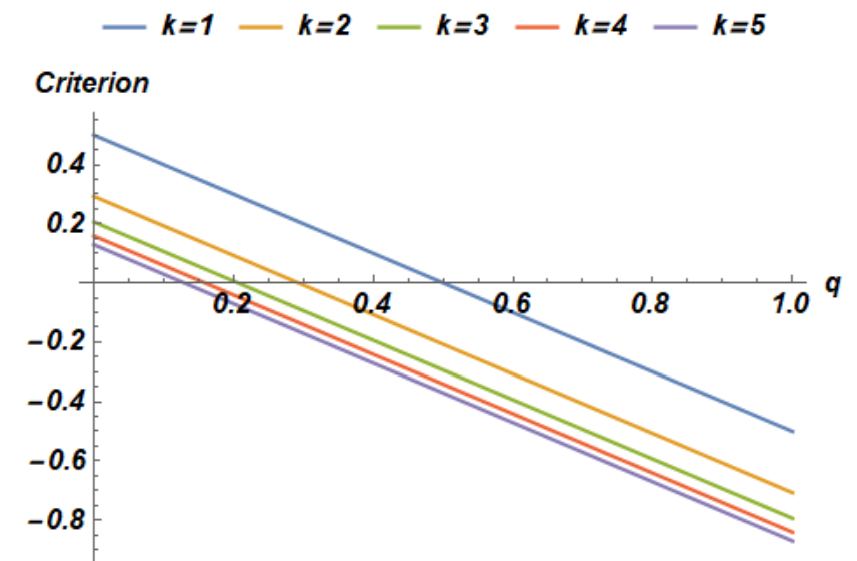}\\
			\caption{\emph{Visualizing the depolarizing channel $\mathcal{N}_{dep}$ as $k$-use non $n$-locality breaking channel for different values of $k.$ Variation of the criterion for $\mathcal{N}_{dep},$ to be non $n$-locality breaking with respect to the parameter characterizing the channel, is plotted. Vertical axis gives the criterion(Eq.(\ref{depo1})) imposed over $q$ for $\mathcal{N}_{dep}$ to be $k$-use non $n$-locality breaking. For any value of $q$ for which the curve lies below the horizontal axis, corresponding channel acts as $k$-use non $n$-locality breaking with $k$ taking some fixed values. No definite conclusion can be given for any portion of the curve above the horizontal axis. }}
			\label{unital1}
		\end{figure}
	\end{center}

	When any depolarizing channel $\mathcal{N}_{dep}$(say) acts over a single-qubit system $\rho,$ the system gets depolarized with probability $q$ and remain unaltered otherwise\cite{nie}:
	\begin{eqnarray}\label{resu2}
		\mathcal{N}_{dep}(\rho)&=& q\frac{\mathbf{I}}{2} +(1-q)\rho.
	\end{eqnarray}
	Alternatively it action can also be expressed as:
	\begin{eqnarray}\label{resu3}
		\mathcal{N}_{dep}(\rho)&=& (1-\frac{3q}{4})\rho+\frac{q}{4}\sum_{i=1}^3\sigma_i\cdot\rho\cdot \sigma_i.
	\end{eqnarray}
	Being unital, $\mathcal{N}_{dep}$ can be written as random unitary channel with equal weights, i.e., can be written in form given by Eq.(\ref{ruc}) for $\alpha,\beta$ given by:
	\begin{eqnarray}
		\alpha&=&\sqrt{1-\frac{3q}{4}}+\mathrm{i}\sqrt{\frac{q}{4}}\nonumber\\
		\beta&=&\sqrt{\frac{q}{4}}(1+\mathrm{i})
	\end{eqnarray}
	So $\mathcal{N}_{dep}$ is one type of single-qubit unital channel. \\
	Clearly, for $\mathcal{N}_{dep},$ we have $Im(\alpha)$$=$$Re(\beta)$$=$$Im(\beta).$ \\
	So $\mathcal{N}_{dep}$ is the special case of $\mathcal{N}_{U}$ as indicated in Theorem.\ref{thec1}.\\
	Let this channel be used $k$ times in $\mathbf{N}_{lin}.$ \\
	By Eq.(\ref{resu1o},) $\mathcal{N}_{dep}$ turns out to be non $n$-locality breaking if:
	\begin{eqnarray}\label{depo1}
		(1-\frac{3q}{4})-(\frac{q}{4})&\leq&\frac{1}{2^{\frac{1}{k}}}\nonumber\\
		\Rightarrow q&\geq&1-2^{-\frac{1}{k}}
	\end{eqnarray}
	Clearly, for any fixed value of noise parameter $q$, the more number of qubits($k$) are passed through the depolarizing channel, lesser is the chance of exhibiting non $n$-locality(see Fig.\ref{unital1}). \\
	For a particular instance, consider $q$$=$$0.4.$ Corresponding $\mathcal{N}_{dep}$ destroys non $n$-locality in case it is used at least $2$ times in $\mathbf{N}_{lin}.$ However, criterion provided by Theorem.\ref{thec1} being sufficient, we cannot give definite conclusion if $\mathbf{N}_{dep}$ is used less than $2$ times.
	
	\subsubsection{\textbf{Dephasing Channel}}
	Let $\mathcal{N}_{dph}$ denote a dephasing channel. $\mathcal{N}_{dph}$ reduces the off diagonal elements of a density operator corresponding to any single-qubit system $\rho$. After action of $\mathcal{N}_{dph},$ $\rho$ loses all superposition and thus behaves as a classical system. Action of $\mathcal{N}_{dph}$ on $\rho$ can be expressed as\cite{nie}:
	\begin{eqnarray}\label{ph1}
		\mathcal{N}_{dph}(\rho)&=& \frac{p}{2}\sigma_3\cdot \rho\cdot \sigma_3 +(1-\frac{p}{2})\rho.
	\end{eqnarray}
	Noise parameter $p$$\in$$[0,1]$ characterizes $\mathcal{N}_{dph}$.
	As $\mathcal{N}_{dph}$ is unital, it can be written as random unitary channel with equal weights(Eq.(\ref{ruc})) for $\alpha,\beta$ now given by:
	\begin{eqnarray}\label{ph2}
		\alpha&=&\sqrt{1-\frac{p}{2}}+\mathrm{i}\sqrt{\frac{p}{2}}\nonumber\\
		\beta&=&0.
	\end{eqnarray}
	Interestingly, $\mathcal{N}_{dph}$ satisfies the restrictions laid down in Theorem.\ref{thec2}. Hence, precisely, we get the following result:
	\begin{corollary}\label{corrc1}
		$\forall k$$=$$1,2,...,2n,$ a single-qubit dephasing channel is not a $k$-use non $n$-locality breaking channel in any linear $n$-local network.
	\end{corollary}
	
	\subsection{Non Unital Channels}
	We consider the class of non unital channels($\mathcal{N}_{NU}$) specified in sec.\ref{prelim} by Eqs.(\ref{nu1},\ref{nu2}). Here the set of parameters characterizing $\mathcal{N}_{NU}$ is given by $\Theta_{\mathcal{N}_{NU}}$$=$$\{t,\lambda_1,\lambda_3\}$ such that Eq.(\ref{nu3}) is satisfied.\\
	Let any such $\mathcal{N}_{NU}$ be used in $\mathbf{N}_{lin}$ such that  $0$$<k$$\leq$$2n$ qubits are passed through $\mathcal{N}_{NU}.$ We next provide the restrictions over the channel parameters for which $\mathcal{N}_{NU}$ turns out to be $k$-use non $n$-locality breaking.
	\begin{theorem}\label{thec3}
		If any non unital channel $\mathcal{N}_{NU}$ from the class specified by Eqs.(\ref{nu1},\ref{nu2}) be such that channel parameters satisfy 
		\begin{eqnarray}\label{resnon1}
			(|t|+|\lambda_3|)^2+\lambda_1^2\leq 1\,\,\textmd{\small{and}}\nonumber\\
			2t^2\cdot \lambda_1^2+\lambda_1^4+(|t|+|\lambda_3|)^4\leq 1\nonumber\\
		\end{eqnarray}
		then $\mathcal{N}_{NU}$ is $k$-use non $n$-locality breaking channel up to $n$-local inequality Eq.(\ref{ineqb}) in any linear $n$-local network.
	\end{theorem}
	\textit{Proof:} See Appendix.\ref{appenc3}.\\
	Any channel from the class of non unital channels $\mathcal{N}_{NU}$ satisfying above conditions(Eq.(\ref{resnon1})) destroys non $n$-local resource when used $k$ times in $\mathbf{N}_{lin}.$ Clearly the observation holds good for any value of $k$ and any length $n$ of the linear chain($\mathbf{N}_{lin}$). 
	\subsubsection{Examples}
\begin{center}
		\begin{figure}
			\includegraphics[width=3.4in]{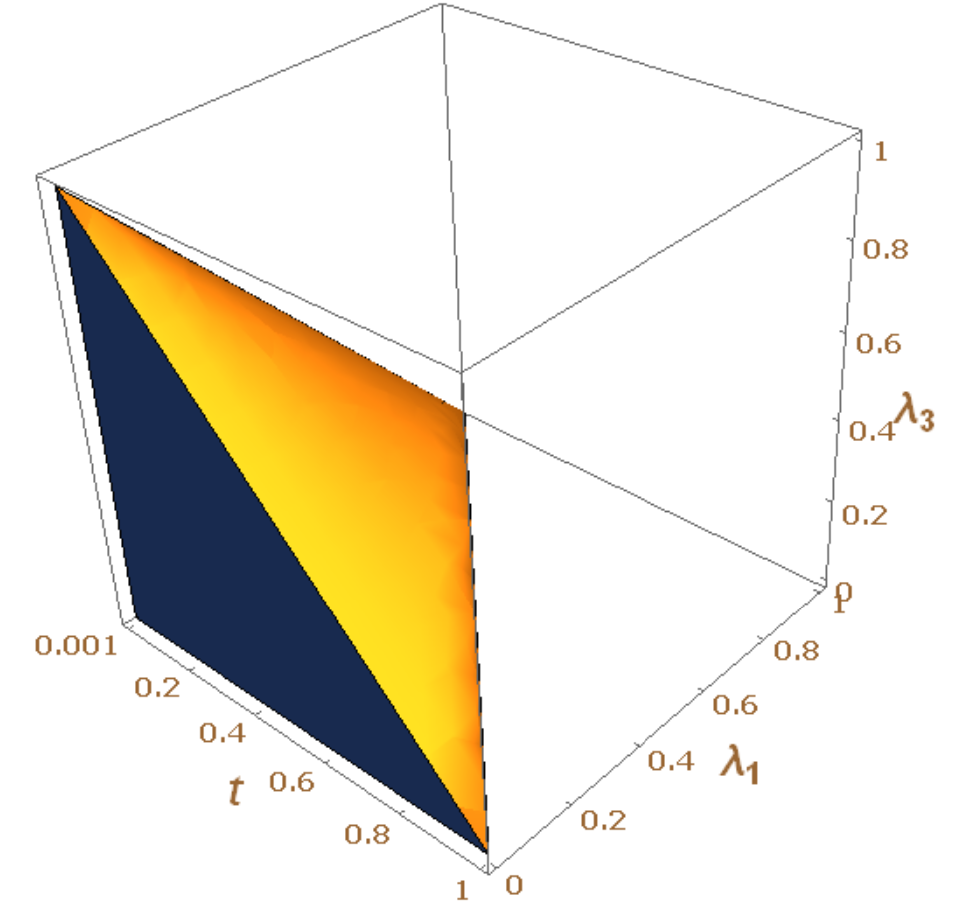}\\
			\caption{\emph{Shaded region gives subspace in the parameter space $(t,\lambda_1,\lambda_3)$ of the class of non unital channels $\mathcal{N}_{NU}$ specified by Eqs.(\ref{nu1},\ref{nu2}). For any tuple of parameter values lying in this region, corresponding non unital channel acts as $k$-use non $n$-locality breaking channel in $\mathbf{N}_{lin}.$}}
			\label{nonfig1}
		\end{figure}
	\end{center}

	Instances of $k$-use non $n$-locality breaking non unital channels(see Fig.\ref{nonfig1}) can be obtained by exploiting conditions laid down in theorem.\ref{thec3} along with the condition specified by Eq.(\ref{nu3}). For a particular example, consider $\mathcal{N}_{NU}$ for $(t,\lambda_1,\lambda_3)$$=$$(0.2,0.2,0.2).$ This channel satisfies the criteria(Eq.(\ref{resnon1})) prescribed by Theorem.\ref{thec3}. So it is $k$-use non $n$-locality breaking channel in $\mathbf{N}_{lin}.$
	Comparing Eq.(\ref{resnon1}) with that of the conditions provided by Eq.(\ref{nu3}) over the channel $\mathcal{N}_{NU},$ it is observed that there exist no region in parameter space for which Eq.(\ref{resnon1}) is violated whereas Eq.(\ref{nu3}) is satisfied. However, a formal proof of the same is beyond the scope of present work. We thus provide the following conjecture:\\
	\textbf{Conjecture.1:} \textit{Any non-unital channel from $\mathcal{N}_{NU}$( Eqs.(\ref{nu1},\ref{nu2})) acts as a non $n$-locality breaking channel in $\mathbf{N}_{lin}.$ }\\

	\section{Characterization in $n$-local Star Network($\textbf{N}_{star}$)}\label{std2}
	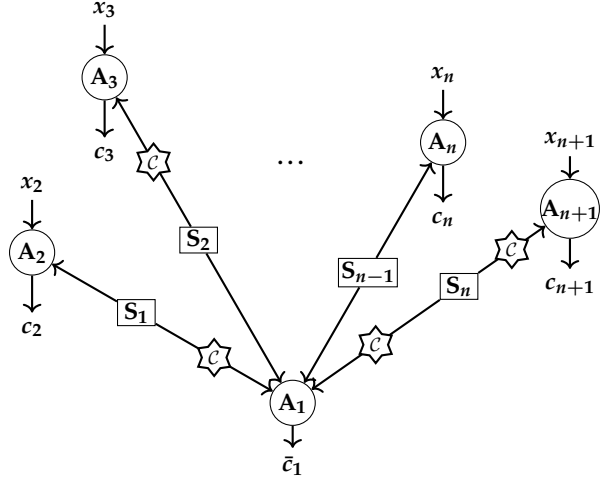
\begin{figure}
		\begin{tikzpicture}[
			node distance=1.4cm and 1.6cm,
			every node/.style={font=\boldmath\bfseries\small},
			meas/.style={
				draw, circle, minimum size=0.6cm,
				inner sep=0pt, align=center,
				text height=1.2ex, text depth=0ex
			},
			source/.style={
				draw, rectangle, minimum width=0.5cm, minimum height=0.3cm,
				inner sep=1pt, align=center
			},
			arr/.style={->, thick},
			chlabel/.style={
				draw, star, star points=6, fill=white,
				inner sep=0.8pt, font=\scriptsize\bfseries
			}
			]
			
			\node[meas] (A1) at (0,0) {$\mathbf{A}_1$};
			
			\node[meas] (A2) at (150:4cm) {$\mathbf{A}_2$};
			\node[meas] (A3) at (120:5cm) {$\mathbf{A}_3$};
			
			\node[meas] (An) at (60:4cm) {$\mathbf{A}_n$};
			\node[meas] (Anp1) at (35:4.5cm) {$\mathbf{A}_{n+1}$};
			
			\node at (90:3.2cm) {\dots};
			
			\node[source] (S1) at ($(A1)!0.6!(A2)$) {$\mathbf{S}_1$};
			\node[source] (S2) at ($(A1)!0.5!(A3)$) {$\mathbf{S}_2$};
			\node[source] (Sn) at ($(A1)!0.5!(An)$) {$\mathbf{S}_{n-1}$};
			\node[source] (Snp1) at ($(A1)!0.6!(Anp1)$) {$\mathbf{S}_n$};
			
			\draw[arr] (S1) -- node[midway,chlabel] {$\mathcal{C}$} (A1);
			\draw[arr] (S1) -- (A2);
			
			\draw[arr] (S2) -- (A1);
			\draw[arr] (S2) -- node[midway,chlabel] {$\mathcal{C}$} (A3);
			\draw[arr] (Sn) -- (A1);
			\draw[arr] (Sn) -- (An);
			\draw[arr] (Snp1)  -- node[midway,chlabel] {$\mathcal{C}$} (A1);
			\draw[arr] (Snp1)  -- node[midway,chlabel] {$\mathcal{C}$} (Anp1);
			\foreach \i/\A in {2/A2, 3/A3, n/An, {n+1}/Anp1} {
				\node at ($(\A)+(0,0.9)$) {$x_{\i}$};
				\draw[arr] ($(\A)+(0,0.7)$) -- (\A);
				\draw[arr] (\A) -- ++(0,-0.8) node[below] {$c_{\i}$};
			}
			
			\draw[arr] (A1) -- ++(0,-0.6) node[below] {$\bar{c}_1$};
		\end{tikzpicture}
		\caption{Single-qubit channel $\mathcal{C}$ acting on some of the qubits in a $n$-local star network $\textbf{N}_{star}.$ }
		\label{nlocalc2}
	\end{figure}
	Let $\varrho_{in}$(Eq.(\ref{tr4})) denote the input state to $\textbf{N}_{star}.$
	Let  an arbitrary single qubit channel $\mathcal{N}$ be used $k$ times in $\textbf{N}_{star}$(see fig.\ref{nlocalc2} for example). As in $\mathbf{N}_{lin},$ here also $k$$\leq$$2n.$ \\
	Let $\mathbf{I}_{star}$ denote the $n$-local inequality(Eq.(\ref{ineqs})) corresponding to $\mathbf{N}_{star}.$ 
	Characterizing $\mathcal{N}$ as a $k$-use non $n$-locality breaking channel in $\textbf{N}_{star}$ will result in imposition of constraints over the set of parameters $\Theta_{\mathcal{N}}$ such that $\mathcal{N}$ destroys non $n$-locality of arbitrary input state $\varrho_{in}$. \\
	We first consider unital channel to explore for such constraints over $\Theta_{\mathcal{N}}.$ 
	\subsection{Unital Channel}
	
	Let an arbitrary random unitary channel $\mathcal{N}_{U}$ be used $k$ times in $\textbf{N}_{star}.$ Here $\Theta_{\mathcal{N}_{U}}$$=$$\{\alpha,\beta\}.$ The restrictions over the channel parameters $\alpha,\beta$ for $\mathcal{N}_{U}$  to be $k$-use non $n$-locality breaking in $\mathbf{N}_{star}$ are formalized in the theorem below.
	\begin{theorem}\label{thec4}
		If any single-qubit unital channel $\mathcal{N}_{U}$ be such that both channel parameters $\alpha,\beta$ are proper complex numbers satisfying:
		\begin{eqnarray}\label{gd1}
			\textmd{Max}[(Re(\alpha))^2,(Re(\beta))^2,(Im(\alpha))^2,(Im(\beta))^2]-\nonumber\\
			\textmd{Min}[(Re(\alpha))^2,(Re(\beta))^2,(Im(\alpha))^2,(Im(\beta))^2]\leq \frac{1}{2^{1+\frac{n}{2k}}},\nonumber\\
		\end{eqnarray}
		then $\mathcal{N}_{U}$ is $k$-use non $n$-locality breaking channel up to $n$-local inequality Eq.(\ref{ineqs}) in any $n$-local star network.\\
		In particular, if $\mathcal{N}_{U}$ be such that any three of  $Re(\alpha),Im(\alpha),Re(\beta),Im(\beta)$ are identical for proper complex numbers $\alpha,\beta$ and satisfy:
		\begin{eqnarray}\label{gd1o}
			\textmd{Max}[(Re(\alpha))^2,(Re(\beta))^2,(Im(\alpha))^2,(Im(\beta))^2]-\nonumber\\
			\textmd{Min}[(Re(\alpha))^2,(Re(\beta))^2,(Im(\alpha))^2,(Im(\beta))^2]\leq \frac{1}{2^{\frac{n}{2k}}},\nonumber\\
		\end{eqnarray} 
		then such $\mathcal{N}_{U}$ is $k$-use non $n$-locality breaking channel up to $n$-local inequality Eq.(\ref{ineqs}) 
	\end{theorem}
	
	\textit{Proof:} Similar to the proof of Theorem.\ref{thec1}.\\
	Theorem.\ref{thec4} gives a sufficient criterion for a single-qubit random unitary channel, hence a unital channel to be non $n$-locality breaking. However the criterion is not necessary as violating above restrictions(Eqs.(\ref{gd1},\ref{gd1o})) leads to no definite conclusion. Unlike Theorem.\ref{thec1}, the restriction imposed over $\alpha,\beta$ now depend not only on how many times $\mathcal{N}_U$ is used in $\mathbf{N}_{star},$ but also on the number of edge parties involved in the network. \\
	Analogous to the linear network scenario, here also the class of $\mathcal{N}_U$ having both $\alpha,\beta$ as improper complex numbers, cannot act as a non $n$-locality breaking channel. We thus formally give the analogue of Theorem.\ref{thec2} in $\mathbf{N}_{star}$:
	\begin{theorem}\label{thec5}
		If any single-qubit unital channel $\mathcal{N}_{U}$ be such that both the channel parameters $\alpha,\beta$ are improper complex numbers or any one of them is null, then $\forall k$$=$$1,2,...,2n,$  $\mathcal{N}_{U}$ is not a $k$-use non $n$-locality breaking channel in any $n$-local star network.
	\end{theorem}
	\textit{Proof:} Similar to that of the proof of Theorem.\ref{thec2}.\\
	Consider a $n$-local star network $\mathbf{N}_{star}$ where any channel from the class of $\mathcal{N}_{U}$ prescribed by Theorem.\ref{thec5} is used $k$ time for any $1$$\leq$$k$$\leq$$2n.$
	Above theorem guarantees existence of collection of arbitrary two-qubit states, which when used in such a network($\mathbf{N}_{star}$), can generate non $n$-local correlations.\\
	Together theorems.\ref{thec2} and \ref{thec5} thus prescribe a class of not $k$-use non $n$-locality breaking single-qubit unital channels.
	
	\subsubsection{\textbf{Few Well-known Unital Channels}}
\begin{center}
		\begin{figure}
			\begin{tabular}{c}
				\subfloat[]{\includegraphics[trim = 0mm 0mm 0mm 0mm,clip,scale=0.48]{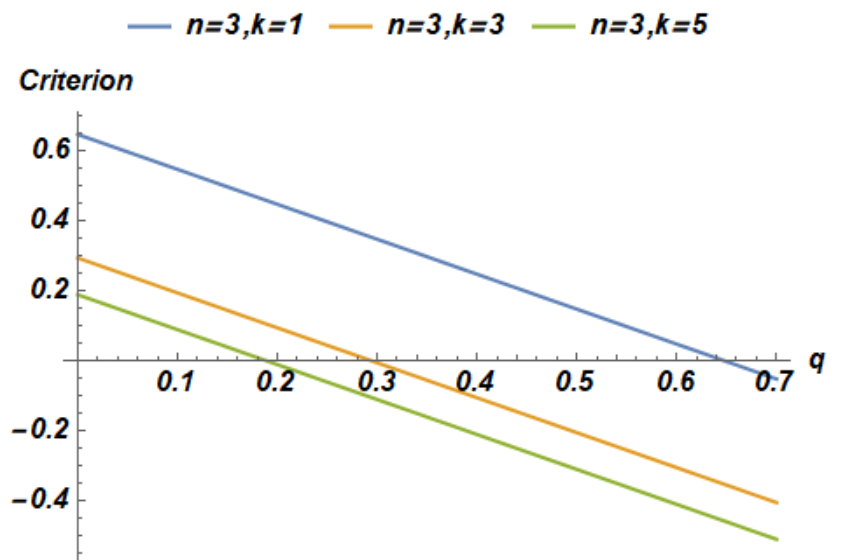}}\\
				\subfloat[]{\includegraphics[trim = 0mm 0mm 0mm 0mm,clip,scale=0.48]{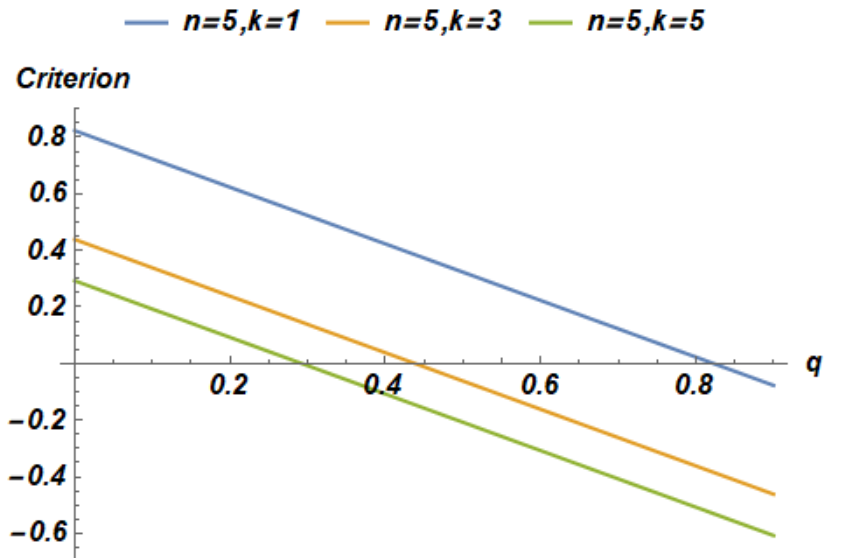}}\\
			\end{tabular}
			\caption{\emph{Both the sub-figures show range of the channel parameter $q$ for which $\mathcal{N}_{dep}$ acts as $k$-use non $n$-locality breaking channel for different values of $k$ in $\mathbf{N}_{star}$ For sub-figure.(a) and (b) we consider different number of independent quantum sources($n$) involved in the network. Variation of the criterion(plotted along vertical axis) for $\mathcal{N}_{dep},$ to be non $n$-locality breaking with respect to the parameter characterizing the channel, is plotted. For any value of $q$ for which the curve lies below the horizontal axis, corresponding channel acts as $k$-use non $n$-locality breaking with $k$ taking some fixed values. No definite conclusion can be given for any portion of the curve above the horizontal axis.}}
			\label{unital2}
		\end{figure}
	\end{center}

	We now analyze implications of above findings for $\mathcal{N}_{dep}$ and $\mathcal{N}_{ph}$.\\
	As we have already seen, for $\mathcal{N}_{dep},$ we have $Im(\alpha)$$=$$Re(\beta)$$=$$Im(\beta).$ \\
	So $\mathcal{N}_{dep}$ is the special case of $\mathcal{N}_{U}$ as indicated in Theorem.\ref{thec4}.\\
	Let this channel be used $k$ times in $\mathbf{N}_{star}.$ \\
	By Eq.(\ref{gd1o},) $\mathcal{N}_{dep}$ turns out to be $k$ use non $n$-locality breaking in $\mathbf{N}_{star}$ if:
	\begin{eqnarray}\label{depos1}
		\Rightarrow q&\geq&1-2^{-\frac{n}{2k}}
	\end{eqnarray}
	Range of noise parameter $q$ for which $\mathcal{N}_{dep}$ destroys non $n$-locality is shown in Fig.\ref{unital2}.

	$\mathcal{N}_{dph}$ belongs to the class of unital channels prescribed in Theorem.\ref{thec4}. Hence $\mathcal{N}_{dph}$ is not a $k$-use non $n$-locality breaking channel in $\mathbf{N}_{star}$.
	\subsection{Non Unital Channel}
	Let any channel from the class of non unital channels $\mathcal{N}_{NU}$(specified by Eqs.(\ref{nu1},\ref{nu2})) be used in $\mathbf{N}_{star}$ such that $0$$<k$$\leq$$2n$ qubits are passed through $\mathcal{N}_{NU}.$  We explore the restrictions over $\lambda_1,\lambda_3,t$(channel parameters) for which $\mathcal{N}_{NU}$ turns out to be $k$-use non $n$-locality breaking.
\begin{theorem}\label{thec6}
	In any $n$-local star network $\mathbf{N}_{star}$, for any fixed tuple of integers $(n,m_1,m_2)$ with $m_1$$+$$2m_2$$=$$k$$\leq$$2n,$ if any non unital channel from $\mathcal{N}_{NU}$ be such that it acts on single subsystem and both subsystems of two-qubit states generated from $m_1$ and $m_2$ sources respectively and the channel parameters satisfy:
	\begin{eqnarray}\label{resnon2}
		(2t^2\cdot \lambda_1^2+\lambda_1^4+(|t|+|\lambda_3|)^4)^{m_2}\cdot
		((|t|+|\lambda_3|)^2+\lambda_1^2)^{m_1}	\leq  2^{\frac{(2-n)(m_1+m_2)}{2}}
	\end{eqnarray}
	then $\mathcal{N}_{NU}$ is $k$-use non $n$-locality breaking channel up to $n$-local inequality Eq.(\ref{ineqs}) in any $n$-local star network.
\end{theorem}
\textit{Proof:} See Appendix.\ref{appenc4}.\\
Any channel from the class of non unital channels $\mathcal{N}_{NU}$ satisfying above conditions(Eq.(\ref{resnon2})) destroys non $n$-local resource when used $k$ times in $\mathbf{N}_{star}.$ Clearly, for any value of $n,$ to decide whether a channel will destroy non $n$-local resource in $\mathbf{N}_{star}$, we need to have information about how the channel is used in the network, i.e., we need to know values of $m_1,m_2.$ Let us now illustrate above theorem in all possible cases. 
\subsubsection{Illustration of Theorem.\ref{thec6}}
\begin{center}
	\begin{figure}
		\begin{tabular}{cc}
			\subfloat[]{\includegraphics[trim = 0mm 0mm 0mm 0mm,clip,scale=0.45]{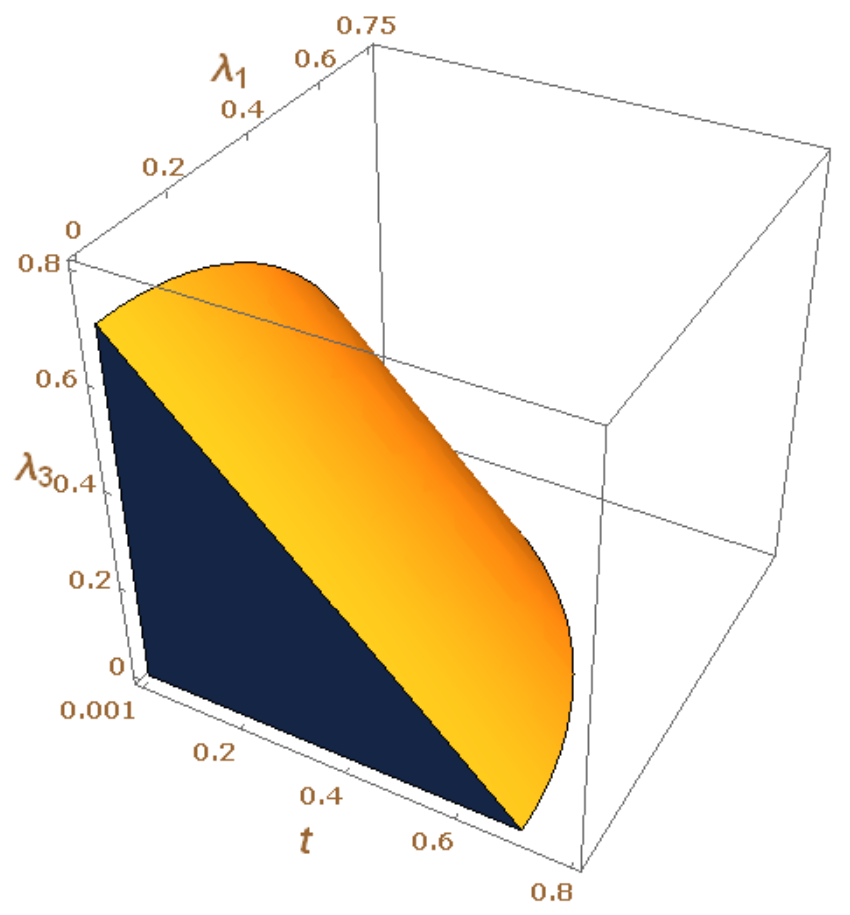}}&{\includegraphics[trim = 0mm 0mm 0mm 0mm,clip,scale=0.45]{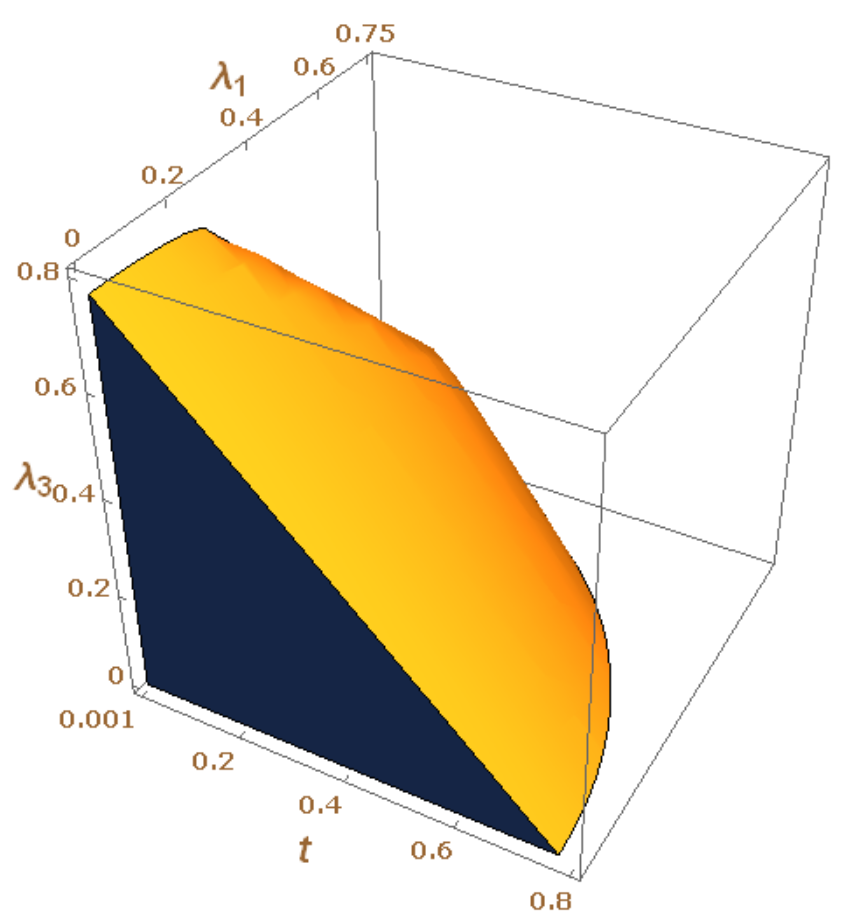}}\\
			\subfloat[]{\includegraphics[trim = 0mm 0mm 0mm 0mm,clip,scale=0.45]{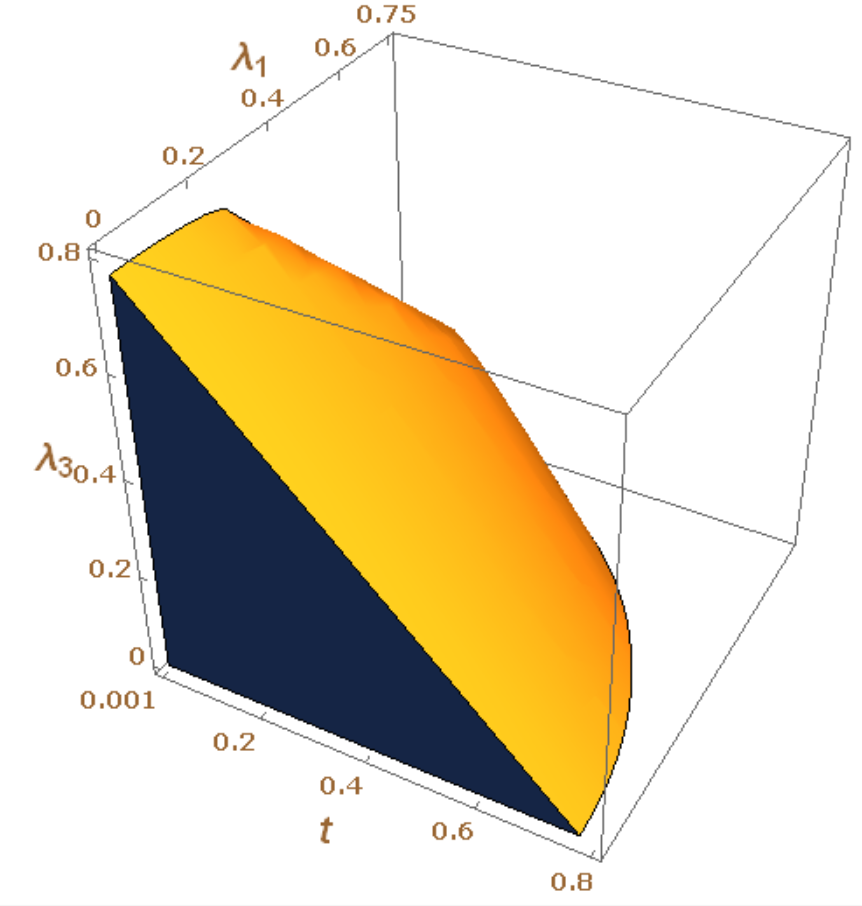}}\\
		\end{tabular}
		\caption{\emph{Each sub-figure provides a subspace in the parameter space $(t,\lambda_1,\lambda_3)$ for non-unital channel from the class $\mathcal{N}_{NU}.$ $n$$=$$4$ for all sub-figures. For sub-fig.(c), we use $(m_1,m_2)$$=$$(2,1)$. For each of these fixed values corresponding sub-figure indicates region in the parameter space for which corresponding member from $\mathcal{N}_{NU}$ acts as $k$-use non $n$-locality breaking channel in $\mathbf{N}_{star}$ having $n$ number of independent quantum sources.}}
		\label{nonfig2}
	\end{figure}
\end{center}

Using theorem.\ref{thec6}, we can find several instances of $k$-use non $n$-locality breaking non unital channels(see sub-figs in Fig.\ref{nonfig2}). We illustrate individually for following cases:
\begin{enumerate}
	\item [1] $\mathcal{N}_{NU}$ acts only on single subsystem of each of the two-qubit state generated by some or all sources: $2n$$\geq$$m_1$$\neq$$0,$ $m_2$$=$$0$.
	\item [2] $\mathcal{N}_{NU}$ acts on both subsystems of each of the two-qubit state generated by some or all sources: $2n$$\geq$$2m_2$$\neq$$0,$ $m_1$$=$$0$.
	\item [3] $\mathcal{N}_{NU}$ acts on single subsystem of each of the two-qubit state generated by some sources and on both subsystems of two-qubit states generated by some other sources: $2n$$\geq$$m_1,2m_2$$\neq$$0.$.
\end{enumerate}
\paragraph{Case.$1$:} Here $k$$=$$m_1.$ The criterion(Eq.(\ref{resnon2})) for any member from $\mathcal{N}_{NU}$ turns out to be: 
\begin{eqnarray}\label{resnon2i}
	(|t|+|\lambda_3|)^2+\lambda_1^2\leq 2^{\frac{2-n}{2}}
\end{eqnarray}
$\mathcal{N}_{NU}$ satisfying Eq.(\ref{resnon2i})  acts as $k$-use non $n$-locality breaking channel. Clearly the criterion depends only on the number($n$) of quantum sources in the network and not on how many two qubit states($k$) each of whose single subsystem come under the action of the channel. So in case Eq.(\ref{resnon2i}) is satisfied for some value of $n,$ then for any $1$$\leq$$k$$\leq$$2n,$ corresponding channel destroys non $n$-locality in $\mathbf{N}_{star}$(see sub-fig.(i) of Fig.\ref{nonfig2} for instance).
\paragraph{Case.$2$:} Here $k$$=$$2m_2.$ The criterion(Eq.(\ref{resnon2})) for any member from $\mathcal{N}_{NU}$ turns out to be:
\begin{eqnarray}\label{resnon2ii}
	2t^2\cdot \lambda_1^2+\lambda_1^4+(|t|+|\lambda_3|)^4	\leq 2^{\frac{2-n}{2}}
\end{eqnarray}
Just as in previous case, here also the criterion is independent of how many two qubit states($m_2$) each of whose both subsystems come under the action of the channel. So in case Eq.(\ref{resnon2ii}) is satisfied for some value of $n,$ then for any $2$$\leq$$k$$\leq$$2n,$ corresponding channel destroys non $n$-locality(see sub-fig.(ii) of Fig.\ref{nonfig2}). 
\paragraph{Case.$3$:} Here $k$$=$$m_1$$+$$2m_2,$ with both $m_1,m_2$$\neq$$0.$  
Unlike previous two cases, here the criterion(Eq.(\ref{resnon2})) involves not only $n$ but also depend on the count of two-qubit states whose single($m_1$) or both($m_2$) subsystem(s) pass through the channel. So, only when for some $(m_1,m_2,n)$ channel parameters satisfy Eq.(\ref{resnon2}), corresponding channel will act as $k$-use non $n$-locality breaking(see sub-fig.(iii) of Fig.\ref{nonfig2}).
	It is clear from Theorem.\ref{thec6} that not every member of $\mathcal{N}_{NU}$ is non $n$-locality breaking channel in $\mathbf{N}_{star}.$ It thus becomes interesting to explore non-unital channels that will not destroy non $n$-local resource for arbitrary quantum sources in $\mathbf{N}_{star}.$ We provide criteria for members from the class $\mathcal{N}_{NU}$ to be resource(non $n$-locality) preserving in $\mathbf{N}_{star}.$
	\begin{theorem}\label{thec7}
		In any $n$-local star network $\mathbf{N}_{star}$, for any fixed tuple of integers $(m_1,m_2)$ with $m_1$$+$$2m_2$$=$$k$$\leq$$2n,$ if any non-unital channel from $\mathcal{N}_{NU}$ be such that it acts on single subsystem and both subsystems of two-qubit states generated from $m_1$ and $m_2$ sources respectively and the channel parameters satisfy:
		\begin{eqnarray}\label{resnon3}
			\sqrt{Y_1^{\frac{m_1}{n}}\cdot Y_3^{\frac{m_2}{n}}+Y_2^{\frac{m_1}{n}}\cdot Y_4^{\frac{m_2}{n}}}>1\nonumber\\
			Y_1=\textmd{\small{Max}}[\lambda_1^2,\lambda_3^2];\,
			Y_2=\textmd{\small{Min}}[\lambda_1^2,\lambda_3^2];\nonumber\\
			Y_3=\textmd{\small{Max}}[\lambda_1^4,(t^2+\lambda_3^2)^2];\nonumber\\
			Y_4=\textmd{\small{Min}}[\lambda_1^4,(t^2+\lambda_3^2)^2];\nonumber\\
		\end{eqnarray}
		then the channel is not a $k$-use non $n$-locality breaking channel in $\mathbf{N}_{star}.$
	\end{theorem}
	\textit{Proof:} See Appendix \ref{appenc5}. \\
	From above theorem, it is clear that for any fixed $n$, $\mathcal{N}_{NU}$ will be non $n$-locality preserving depending not only on the number of qubits getting distributed through $\mathcal{N}_{NU},$ but also on the pattern of distribution. Precisely whether single or both subsystems of a two-qubit state coming under action of the channel. Let us use numerical instances to illustrate the three possible cases involving theorem.\ref{thec7}.
	\subsubsection{Illustration of Theorem.\ref{thec7}}
\begin{center}
		\begin{figure}
			\begin{tabular}{cc}
				\subfloat[]{\includegraphics[trim = 0mm 0mm 0mm 0mm,clip,scale=0.45]{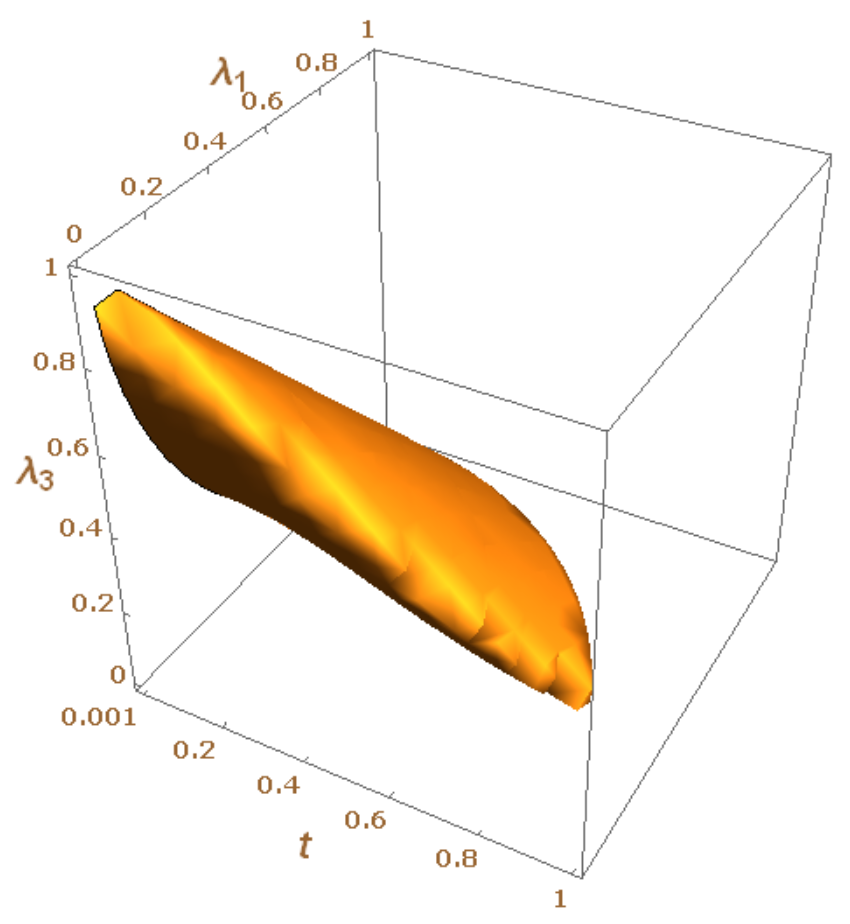}}&{\includegraphics[trim = 0mm 0mm 0mm 0mm,clip,scale=0.45]{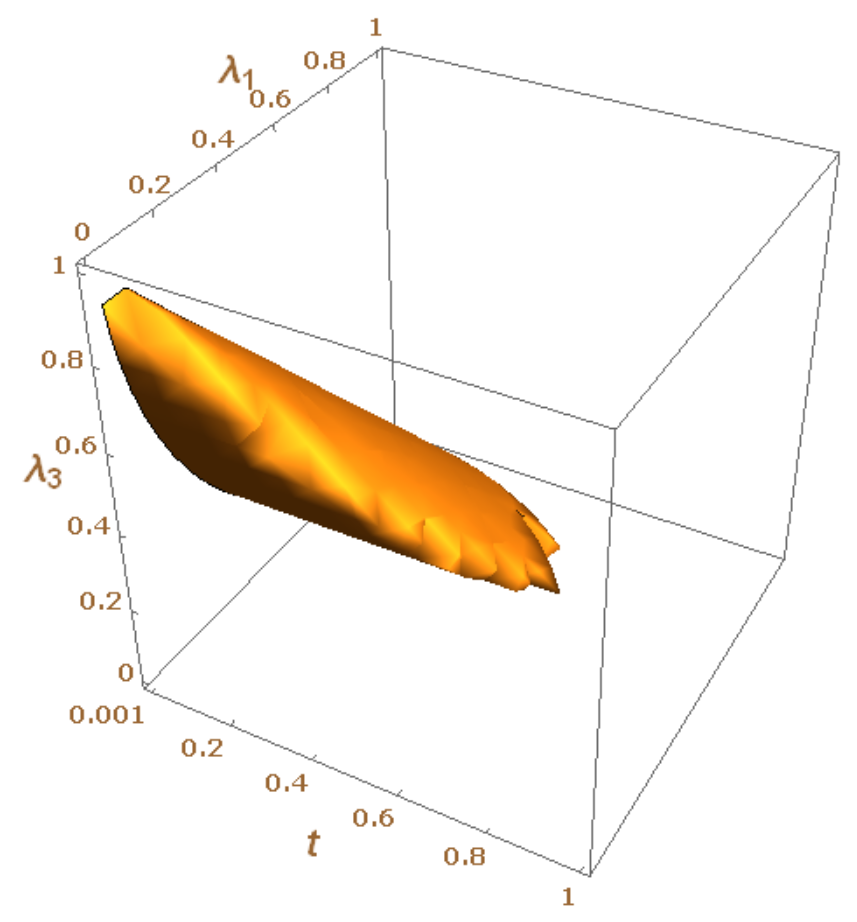}}\\
				\subfloat[]{\includegraphics[trim = 0mm 0mm 0mm 0mm,clip,scale=0.45]{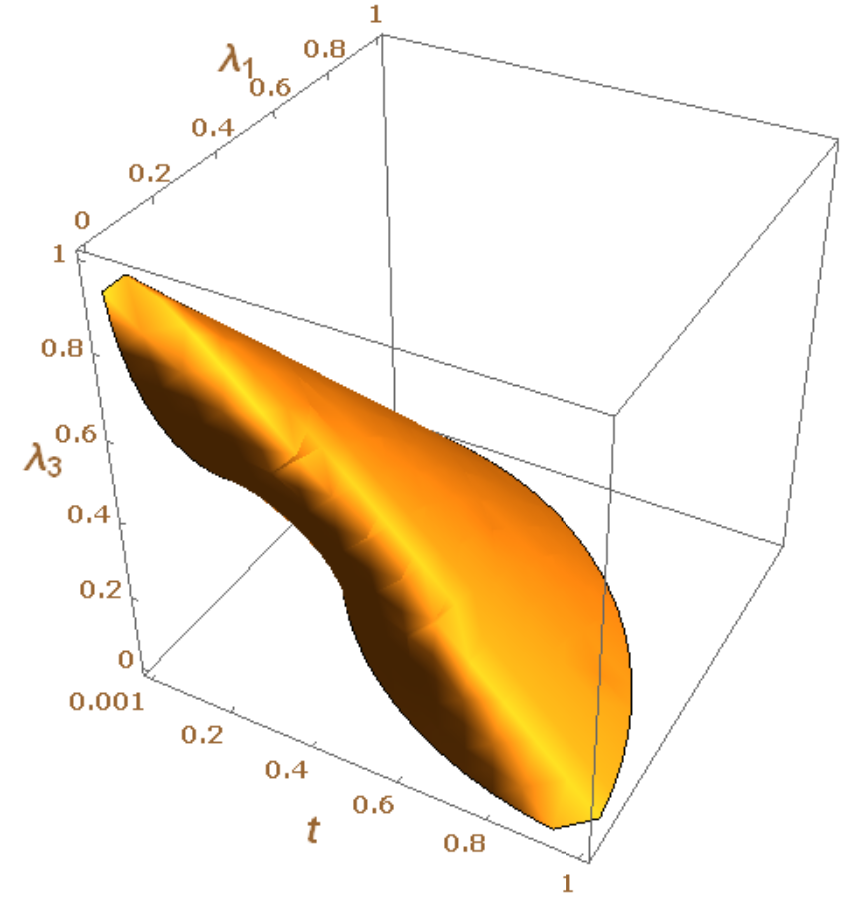}}\\
			\end{tabular}
			\caption{\emph{Each sub-figure provides a subspace in the parameter space $(t,\lambda_1,\lambda_3)$ for non-unital channel from the class $\mathcal{N}_{NU}.$ The specifications of $(m_1,m_2,n)$ for sub-figures.(a),(b),(c) are $(4,0,14),$ $(0,2,14)$ and $(2,1,14)$ respectively. For each of these fixed values of $(m_1,m_2,n),$ corresponding sub-figure gives $k$-use non $n$-locality preserving non unital channel in $\mathbf{N}_{star}$. Regions are obtained by using Theorem.\ref{thec7} along with the channel criteria(Eq.(\ref{nu3})).}}
			\label{fal1}
		\end{figure}
	\end{center}

	We have the same three cases as we got while illustrating theorem.\ref{thec6}.
	\paragraph{Case.$1$:} Let us consider a $14$-local star network, i.e., $n$$=$$14.$ W.L.O.G., let a non unital channel from $\mathcal{N}_{NU}$ acts only on single subsystem of each of the two-qubit state generated by sources $\mathbf{S}_1,\mathbf{S}_2\mathbf{S}_3,\mathbf{S}_4.$ While rest of the qubits do not come under the action of the channel in $\mathbf{N}_{star}.$ So, here $m_1$$=$$4$ and $m_2$$=$$0$ and hence $k$$=$$4.$ Several members from $\mathcal{N}_{NU}$ do not act as $4$-use non $14$-locality breaking channel in the network(see sub-fig.(a) of Fig.\ref{fal1}). 
	\paragraph{Case.$2$:} Let us consider the scenario almost similar to case.$1.$ Except now let a non-unital channel from $\mathcal{N}_{NU}$ acts on both subsystems of each of the two-qubit state generated by sources $\mathbf{S}_1,\mathbf{S}_2.$ So, here $m_1$$=$$0$ and $m_2$$=$$2$ and hence $k$$=$$4.$ There exist non-unital channels from $\mathcal{N}_{NU}$ that do not act as $4$-use non $14$-locality breaking channel in the network(see sub-fig.(b) of Fig.\ref{fal1}). 
	\paragraph{Case.$3$:} Lastly consider $n$$=$$14,$ $(m_1,m_2)$$=$$(2,1).$ So, now let us consider channel acts on single subsystem of each of the two-qubit state generated by two sources($\mathbf{S}_1,\mathbf{S}_2,$say) and acts on both subsystems of a two-qubit state coming from another source($\mathbf{S}_3$,say). There exist instances of channels not destroying non $14$-locality in this case also(see sub-fig.(c) in Fig.\ref{fal1}).
	So far we have considered standard $n$-local networks where all sources may not be non-local. We next provide our observations regarding non standard $n$-local networks.
	\section{Characterizing Channels from Perspective of Full Network Nonlocality}\label{fulln}
	Full network correlations turn out to be a stronger notion of non $n$-local correlations. As already mentioned in sec.\ref{prelim}, there exist upper bound(Eq.(\ref{up212})) of Bell-type inequality(Eq.(\ref{ineqsn})) detecting full network nonlocality in non standard $3$-local star network only. So, we next consider that particular configuration of non standard network for further discussion. 
	\subsection{$k$-Use Full Network Nonlocality Breaking Channel}
\begin{center}
		\begin{figure}
			\begin{tabular}{cc}
				\subfloat[]{\includegraphics[trim = 0mm 0mm 0mm 0mm,clip,scale=0.45]{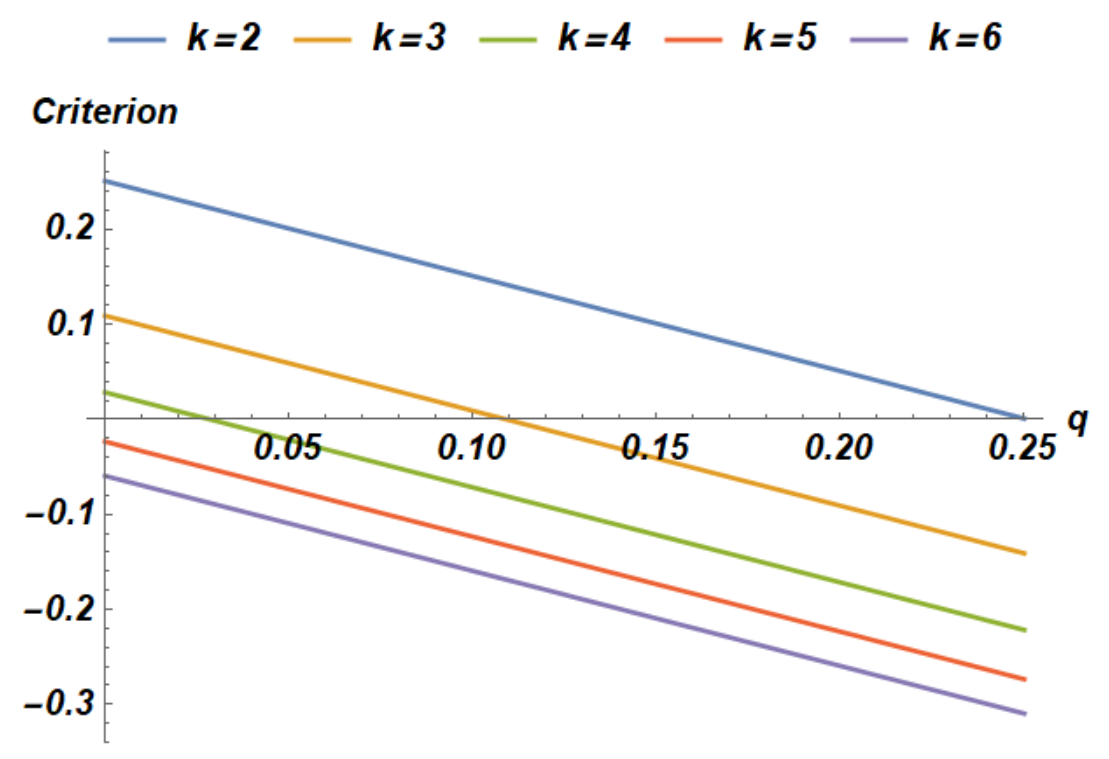}}\\
				\subfloat[]	{\includegraphics[trim = 0mm 0mm 0mm 0mm,clip,scale=0.45]{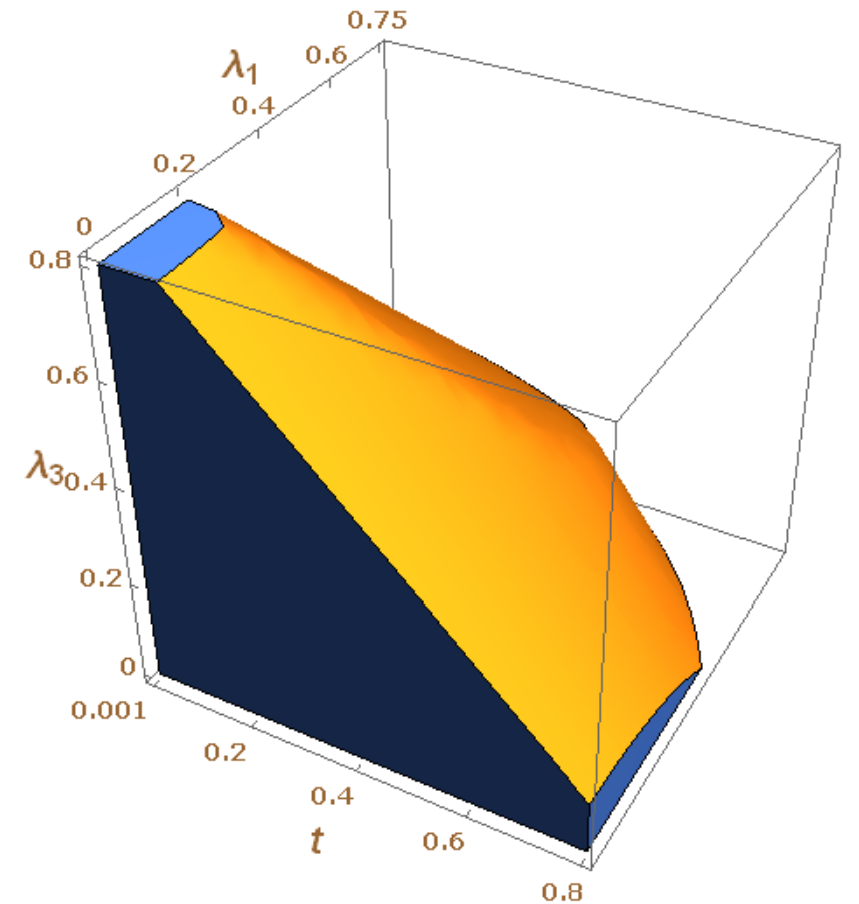}}\\
			\end{tabular}
			\caption{\emph{Visualizing the role of depolarization channel(sub-fig.(a)) and that of $\mathcal{N}_{NU}$(sub-fig.(b)) in destroying full network nonlocality. In sub-fig.(a) for each possible value of $k$$\geq$$5,$ straight line corresponding to the criterion provided by Eq.(\ref{deposf1}) lies entirely below $q$ axis. This indicates any $\mathcal{N}_{dep}$ acts as $k$-use full network nonlocality breaking channel when used at least five times in the network. In sub-fig.(b) we have considered $(m_1,m_2)$$=$$(2,1).$ Corresponding to any point lying in the shaded region $\mathcal{N}_{NU}$ destroys the same resource.}}
			\label{full1}
		\end{figure}
	\end{center}

	We formalize our observations for a unital channel in the following theorem.
	\begin{theorem}\label{thec8}
		If any single-qubit unital channel $\mathcal{N}_{U}$ be such that both channel parameters $\alpha,\beta$ are proper complex numbers satisfying:
		\begin{eqnarray}\label{gdf1}
			\textmd{Max}[(Re(\alpha))^2,(Re(\beta))^2,(Im(\alpha))^2,(Im(\beta))^2]-\nonumber\\
			\textmd{Min}[(Re(\alpha))^2,(Re(\beta))^2,(Im(\alpha))^2,(Im(\beta))^2]\leq \frac{1}{2^{\frac{4k+9}{6k}}},\nonumber\\
		\end{eqnarray}
		then $\mathcal{N}_{U}$ is $k$-use full network nonlocality breaking channel up to trilocal inequality Eq.(\ref{ineqsn}) in any non standard trilocal star network.\\
		In particular, if $\mathcal{N}_{U}$ be such that any three of  $Re(\alpha),Im(\alpha),Re(\beta),Im(\beta)$ are identical for proper complex numbers $\alpha,\beta$ and satisfy:
		\begin{eqnarray}\label{gdf1o}
			\textmd{Max}[(Re(\alpha))^2,(Re(\beta))^2,(Im(\alpha))^2,(Im(\beta))^2]-\nonumber\\
			\textmd{Min}[(Re(\alpha))^2,(Re(\beta))^2,(Im(\alpha))^2,(Im(\beta))^2]\leq \frac{1}{2^{\frac{9-2k}{6k}}},\nonumber\\
		\end{eqnarray} 
		then such $\mathcal{N}_{U}$ is $k$-use full network nonlocality breaking channel with respect to Eq.(\ref{ineqsn}) in non standard trilocal star network.
	\end{theorem}
	\textit{Proof:} Similar to the proof of Theorem.\ref{thec1}.\\
	By criterion(Eq.(\ref{gdf1o})), depolarizing channel($\mathcal{N}_{dep}$) destroys full network nonlocality if
	\begin{eqnarray}\label{deposf1}
		q&\geq&1- 2^{\frac{2k-9}{6k}}
	\end{eqnarray}
	Exploiting this criterion(Eq.(\ref{deposf1})), it can be easily seen that when a depolarizing channel is used at least $3$ times in the network, full network nonlocality is not detected by Eq.(\ref{ineqsn}) for any value of the noise parameter $q$(see  sub-fig.(a) in Fig.\ref{full1}).\\
	We now state the observations related to non unital channels through the following theorem,
	\begin{theorem}\label{thec9}
		In any non standard trilocal star network $\mathbf{N}_{star}$, for any fixed tuple of integers $(m_1,m_2)$ with $m_1$$+$$2m_2$$=$$k$$\leq$$6,$ if any non unital channel from $\mathcal{N}_{NU}$ be such that it acts on single subsystem and both subsystems of two-qubit states generated from $m_1$ and $m_2$ sources respectively and the channel parameters satisfy:
		\begin{eqnarray}\label{resnonf2}
			(2t^2\cdot \lambda_1^2+\lambda_1^4+(|t|+|\lambda_3|)^4)^{m_2}\cdot
			((|t|+|\lambda_3|)^2+\lambda_1^2)^{m_1}	\leq 2^{\frac{2-3(m_1+m_2)}{6}}
		\end{eqnarray}
		then $\mathcal{N}_{NU}$ is $k$-use full network nonlocality breaking channel with respect to trilocal inequality Eq.(\ref{ineqsn}) in the network.
	\end{theorem}
	\textit{Proof:} Similar to the proof of Theorem.\ref{thec6}.\\
	There exist members from $\mathcal{N}_{NU}$ which can destroy full network nonlocality(see sub-fig.(ii) of Fig.\ref{full1}).
	\vskip 10cm

	\section{Conclusions and Discussions} \label{conc}

     Evolution of quantum states under effect of environmental noises is usually described by action of channels on the states. Present work involves study of the action of single-qubit channels in destroying resource of network non locality in both linear and non-linear(star topology) $n$-local networks. Results presented here characterize both unital and a class of non-unital channels(Eq.(\ref{nu2})) as non $n$-locality breaking channels for any finite number of sources($n$). Interestingly, present work has also prescribed sufficient conditions, which when satisfied, corresponding unital or some non-unital channels(Eq.(\ref{nu2})) may destroy a resource as strong as full network nonlocality. To this end, it may be noted that presence of any form of network nonlocality, considered in this work, depends upon its detection via violation of suitable $n$-local inequality. However, violation of any such inequality is only a sufficient nonlocality detection criterion. So any set of network correlations, obtained after a channel's action($\mathcal{N}$,say), may still be inexplicable in terms of any $n$-local hidden variable model even if it fails to show $n$-local inequality violation. However, ensuring nonlocality via construction of such models is complicated for obvious reasons and also corresponding nonlocal correlations may not be detected in any practical experimental set-up.
So, to point out whether a channel is network nonlocality breaking, the present work relies only on practically detectable form of network nonlocality obtained via violation of suitable $n$-local inequalities . However, considering explicit handling of $n$-local models in this context is indeed a potential direction of future research.

Any channel that preserves network nonlocal resource is important for practical tasks. Present work has been able to point out some such channels. More importantly, for some  single-qubit channels, this work prescribes network non $n$-locality preserving criteria. So, in totality, present study provides criteria both for destruction and preservation of network nonlocal correlations under action of channels. However, there still exist several single-qubit channels that do not satisfy any of these criteria. Complete characterization of the same warrants further investigation.

As a byproduct of the study, role of two noisy channels such as depolarizing and phase damping, has been analyzed. It will be interesting to extend the study further so that a systematic analysis of larger class of noisy channels can be provided.

\appendix
    
	\section{}\label{appenc1} 
	Here we give proof of Theorem.\ref{thec1}.\\
	Given that action of $\mathcal{N}_{U}$ is equivalent to that of a random unitary channel with equal weights. Action of $\mathcal{N}_{U}$ on any single qubit state $\rho$ thus can be written as $\mathcal{N}_{U}(\rho)$$=$$\frac{1}{4}\sum_{i=1}^4 U_i\cdot \rho\cdot U_i^{\dagger}$ where $U_1,U_2,U_3,U_4$ are all unitary operations satisfying \cite{choiunital}:
	\begin{eqnarray}\label{appa1}
		U_1 &=& \begin{bmatrix}
			\alpha & \bar{\beta} \\
			-\beta & \bar{\alpha} 
		\end{bmatrix} \nonumber\\
		U_2& =& \begin{bmatrix}
			\alpha & -\bar{\beta} \\
			\beta & \bar{\alpha} 
		\end{bmatrix} \nonumber\\
		U_3 &=& \begin{bmatrix}
			\bar{\alpha} & \beta \\
			-\bar{\beta} & \alpha 
		\end{bmatrix} \nonumber\\
		U_4 &=& \begin{bmatrix}
			\bar{\alpha} & -\beta \\
			\bar{\beta} & \alpha 
		\end{bmatrix} \\
		\alpha, \beta& \in& \mathbb{C}\,\textmd{ with } |\alpha|^2 + |\beta|^2 = 1
	\end{eqnarray}
	Effect of $\mathcal{N}_{U}$ on the Pauli matrices are:
	\begin{eqnarray}\label{appa2}
		\mathcal{N}(\sigma_0)&=&\sigma_0\nonumber\\
		\mathcal{N}(\sigma_1)&=&\frac{\alpha^2+\bar{\alpha}^2-\beta^2-\bar{\beta}^2}{2}\sigma_1\nonumber\\
		\mathcal{N}(\sigma_2)&=&\frac{\alpha^2+\bar{\alpha}^2+\beta^2+\bar{\beta}^2}{2}\sigma_2\nonumber\\
		\mathcal{N}(\sigma_3)&=&(|\alpha|^2-|\beta|^2)\sigma_3\\
	\end{eqnarray}
	Let $\varrho_i$ denote arbitrary two-qubit state with correlation tensor $\mathbf{W}_i$
	Let us first see the change in eigen values of $\textbf{W}_i$$=$$\textmd{diag}(w_{i1},w_{i2},w_{i3})$ due to action of $\mathcal{N}.$ For that let $\textbf{W}^{'}_i$$=$$\textmd{diag}(w^{'}_{i1},w^{'}_{i2},w^{'}_{i3})$ denote the correlation tensor of $\varrho_i^{'}$(resulting from action of $\mathcal{N}_{U}$).\\
	Now depending on whether single or both qubits of $\varrho_i$ are acted upon by $\mathcal{N}_{U}$ we have following two possibilities:
	\begin{enumerate}
		\item [1.] \textit{When $\mathcal{N}_{U}$ acts on one of two qubits(say, $1^{st}$ qubit) of $\varrho_i:$}Here, $\varrho_i^{'}$$=$$\mathcal{N}\otimes id_2(\varrho_i).$ \\
		Using Eq.\ref{appa2}, we get:
		\begin{eqnarray}\label{appa3}
			w^{'}_{i1}&=&\frac{\alpha^2+\bar{\alpha}^2-\beta^2-\bar{\beta}^2}{2}w_{i1}\nonumber\\
			w^{'}_{i2}&=&\frac{\alpha^2+\bar{\alpha}^2+\beta^2+\bar{\beta}^2}{2}w_{i2}\nonumber\\
			w^{'}_{i3}&=&(|\alpha|^2-|\beta|^2)w_{i3}
		\end{eqnarray}
		\item [2.] \textit{When $\mathcal{N}_{U}$ acts on both the qubits of $\varrho_i:$}Here, $\varrho_i^{'}$$=$$\mathcal{N}\otimes id_2(\varrho_i).$ \\
		Using Eq.\ref{appa2}, we get:
		\begin{eqnarray}\label{appa4}
			w^{'}_{i1}&=&\frac{(\alpha^2+\bar{\alpha}^2-\beta^2-\bar{\beta}^2)^2}{4}w_{i1}\nonumber\\
			w^{'}_{i2}&=&\frac{(\alpha^2+\bar{\alpha}^2+\beta^2+\bar{\beta}^2)^2}{4}w_{i2}\nonumber\\
			w^{'}_{i3}&=&((|\alpha|^2-|\beta|^2)^2)w_{i3}
		\end{eqnarray}
	\end{enumerate}
	$\mathcal{N}_{U}$ is used $k$ times in the network.
	As discussed in the main text, depending on which of $2n$ are acted upon by $\mathcal{N}_{U}$, we can have different cases(see the list in \ref{enu1}). Combining all the possibilities Case.(iii) gave the most general possibility. For ease of discussion, let us recall back the most general situation: \\
	W.L.O.G., let $\mathcal{N}_{U}$ act upon single qubit of each $\varrho_1,\varrho_2,...,\varrho_{m_1}$ and both qubits of each of $\varrho_{m_1+1},\varrho_{m_1+2},...,\varrho_{m_1+m_2}$ whereas does not act upon any qubit of remaining $n$$-$$m_1$$-$$m_2$ states $\varrho_{m_1+m_2+1},$$\varrho_{m_1+m_2+2},$...$\varrho_{n}.$\\
	So, we have:
	\begin{eqnarray}\label{k-val}
		k=m_1+2m_2
	\end{eqnarray}
	The overall state in the network $\varrho_{in}^{'}$ is given by Eq.(\ref{nb4}), i.e.,
	\begin{eqnarray*}
		\varrho_{in}^{'}&=&\otimes_{j=1}^{m_1}(\mathcal{N}\otimes id(\varrho_j)\otimes_{j=m_1+1}^{m_1+m_2}(\mathcal{N}\nonumber\\
		&&\otimes \mathcal{N}(\varrho_j))\otimes_{j=m_1+m_2+1}^n\varrho_j
	\end{eqnarray*}
	Using Eq.(\ref{appa3}), $\forall i$$=$$1,2,...,m_1,$ we get the following form of correlation tensor $\textbf{W}_i^{'}$ of $\varrho_i^{'}:$
	\begin{eqnarray}\label{appa5}
		\textbf{W}_i^{'}&=&\textmd{diag}(w_{i1}^{'},w_{i2}^{'},w_{i3}^{})\,\,\textmd{\small{where}}\nonumber\\
		w^{'}_{i1}&=&\frac{|\alpha^2+\bar{\alpha}^2-\beta^2-\bar{\beta}^2|}{2}w_{i1}\nonumber\\
		w^{'}_{i2}&=&\frac{|\alpha^2+\bar{\alpha}^2+\beta^2+\bar{\beta}^2|}{2}w_{i2}\nonumber\\
		w^{'}_{i3}&=&(||\alpha|^2-|\beta|^2|)w_{i3}\,\forall i=1,...,m_1\nonumber\\
		&&
	\end{eqnarray}
	Using Eq.(\ref{appa4}), $\forall i$$=$$m_1$$+$$1,m_1$$+$$2,$...,$m_1$$+$$m_2$ we get the following form of correlation tensor $\textbf{W}_i^{'}$ of $\varrho_i^{'}:$
	\begin{eqnarray}\label{appa6}
		\textbf{W}_i^{'}&=&\textmd{diag}(w_{i1}^{'},w_{i2}^{'},w_{i3}^{})\,\,\textmd{\small{where}}\nonumber\\
		w^{'}_{i1}&=&\frac{(|\alpha^2+\bar{\alpha}^2-\beta^2-\bar{\beta}^2|)^2}{4}w_{i1}\nonumber\\
		w^{'}_{i2}&=&\frac{(|\alpha^2+\bar{\alpha}^2+\beta^2+\bar{\beta}^2|)^2}{4}w_{i2}\nonumber\\
		w^{'}_{i3}&=&(||\alpha|^2-|\beta|^2|)^2 w_{i3}\,\forall i=m_1+1,...,m_1+m_2\nonumber\\
		&&
	\end{eqnarray}
	$\forall i$$=$$m_1$$+$$m_2$$+$$1,$...$n,$  correlation tensor $\textbf{W}_i^{'}$ of $\varrho_i$ remains unaltered:
	\begin{eqnarray}\label{appa7}
		\textbf{W}_i^{'}&=&\textmd{diag}(w_{i1}^{'},w_{i2}^{'},w_{i3}^{})\,\,\textmd{\small{where}}\nonumber\\
		w^{'}_{i1}&=&w_{i1}\nonumber\\
		w^{'}_{i2}&=&w_{i2}\nonumber\\
		w^{'}_{i3}&=& w_{i3}\,\forall i=m_1+m_2+1,...,n\nonumber\\
		&&
	\end{eqnarray}
	$\forall i$$=$$1,2,...,n$ let $E_{i1}$$\geq$$E_{i2}$$\geq$$E_{i3}$ denote the ordered singular values of $\textbf{W}_i.$ So  for $j$$=$$1,2,3,$ $E_{ij}$$\in$$\{w_{i1},w_{i2},w_{i3}\}$ \\
	Similarly, $\forall i$$=$$1,2,...,n$ let $E_{i1}^{'}$$\geq$$E_{i2}^{'}$$\geq$$E_{i3}^{'}$ denote the ordered singular values of $\textbf{W}_i^{'}.$ So for $j$$=$$1,2,3,$ $E_{ij}^{'}$$\in$$\{w_{i1}^{'},w_{i2}^{'},w_{i3}^{'}\}$ \\
	Let us introduce three terms $s_1,s_2,s_3$ such that:
	\begin{eqnarray}\label{appa8}
		s_1&=& \frac{|\alpha^2+\bar{\alpha}^2-\beta^2-\bar{\beta}^2|}{2}\nonumber\\
		&=& |(Re(\alpha))^2-(Im(\alpha))^2-(Re(\beta))^2+(Im(\beta))^2|\nonumber\\
		s_2&=&\frac{|\alpha^2+\bar{\alpha}^2+\beta^2+\bar{\beta}^2|}{2}\nonumber\\
		&=& |(Re(\alpha))^2-(Im(\alpha))^2+(Re(\beta))^2-(Im(\beta))^2|\nonumber\\
		s_3&=&||\alpha|^2-|\beta|^2|\nonumber\\
		&=& |(Re(\alpha))^2+(Im(\alpha))^2-(Re(\beta))^2-(Im(\beta))^2|\nonumber\\
		&&
	\end{eqnarray}
	Considering all possibilities provided by Eq.(\ref{appa8}), we get:
	\begin{eqnarray}\label{appa8o}
		s_1,s_2,s_3&\in& \{\sum_{k=1}^4(-1)^{i_k}M_k\}_{i_1,i_2,i_3,i_4},\,\,\textmd{such that }\nonumber\\
		i_1,i_2,i_3,i_4&\in&\{0,1\};\, \sum_{k=1}^4 i_k=2\textmd{ and} \nonumber\\
		M_1=(Re(\alpha))^2&;& M_2=(Im(\alpha))^2\nonumber\\
		M_3=(Re(\beta))^2&;&M_4=(Im(\beta))^2\nonumber\\
		&&
	\end{eqnarray}
	Let
	\begin{eqnarray}\label{opts1}
		M_{max}&=&\textmd{Max}[M_1,M_2,M_3,M_4]\nonumber\\
		M_{Min}&=&\textmd{Min}[M_1,M_2,M_3,M_4].\nonumber\\
		&&
	\end{eqnarray}
	By Eqs.(\ref{appa8o},\ref{opts1}), we have:
	\begin{eqnarray}\label{opts2}
		s_1,s_2,s_3&\leq&2(M_{max}-M_{min})\nonumber\\
		\textmd{Hence,}\,\textmd{Max}_i s_i&\leq&2(M_{max}-M_{min})
	\end{eqnarray}
	We now test $n$-local inequality(Eq.(\ref{ineqb})) using $\varrho_{in}^{'},$ i.e., using $\varrho_{1}^{'},$ $\varrho_{2}^{'},$....,$\varrho_{n}^{'}.$ \\
	Upper bound($\textbf{B}_{n-lin}$) of Eq.(\ref{ineqb}) is given by:
	\begin{eqnarray}\label{appa10}
		\textbf{B}_{n-lin}&=&\sqrt{\Pi_{i=1}^n E_{i1}^{'}+\Pi_{i=1}^n E_{i2}^{'}}\nonumber\\
		&=& \sqrt{\Pi_{r=1}^3s_r^{j_{r}}\Pi_{j=1}^n E_{j1}+ \Pi_{r^{'}=1}^3s_{r^{'}}^{j_{r^{'}}}\Pi_{j=1}^n E_{j2}}\,\,\textmd{where}\nonumber\\
		j_1,j_2,j_3&\in&\{0,1,...,k\}\textmd{s.t }\sum_{i=1}^3j_i=k\,(\textmd{\small{By Eq.(\ref{k-val})}})\nonumber
		\\
		&\leq& \sqrt{ \Pi_{r=1}^3s_r^{j_{r}}+ \Pi_{r^{'}=1}^3s_{r^{'}}^{j_{r^{'}}}}\nonumber\\
		&\leq&\sqrt{2(\textmd{Max}_{i=1}^3 s_i)^k}\nonumber\\
		&\leq&\sqrt{2^{k+1}(M_{max}-M_{min})^k}\,\,\textmd{\small{By Eq.(\ref{opts2})}}\nonumber\\
		&\leq& \sqrt{2^{k+1}(2^{-\frac{k+1}{k}})^k}\,\,\textmd{\small{By Eq.(\ref{resu1})}}\nonumber\\
		&=& 1
	\end{eqnarray}
	This indicates that $\varrho_{in}^{'}$ is not detected as a non $n$-local resource by the $n$-local inequality. \\
	$\varrho_1,\varrho_2,...,\varrho_n$ being arbitrary, $\mathcal{N}_{U}$ turns out to be $k$-use non $n$-locality breaking channel.\\
	Now let us prove the second part of the theorem where we consider the channel parameters to be such that any three of $Re(\alpha),Im(\alpha),Re(\beta),Im(\beta)$ are identical. For such $\mathcal{N}_{RU},$ any three of $M_1,M_2,M_3,M_4$(Eq.(\ref{appa8o})) are identical. \\
	W.L.O.G., let $M_1$$=$$M_2$$=M_3.$ Then Eq.(\ref{appa8o}) gets modified:
	\begin{eqnarray}\label{opts3}
		s_1,s_2,s_3&=&\{M_1-M_4,M_4-M_1\}
	\end{eqnarray}
	Consequently, here we get more stringent restriction over $s_1,s_2,s_3$ than that provided by Eq.(\ref{opts2}):
	\begin{eqnarray}\label{opts4}
		s_1,s_2,s_3&\leq&M_{max}-M_{min}\nonumber\\
		\textmd{Hence,}\,\textmd{Max}_i s_i&\leq&M_{max}-M_{min}
	\end{eqnarray}
	Then approaching as before and using Eq.(\ref{opts4}) instead of Eq.(\ref{opts2}) and condition provided in the theorem by Eq.(\ref{resu1o}), we get the required result:
	\begin{equation}
		\mathbf{B}_{n-lin}\leq 1
	\end{equation}
	This completes the theorem.$\blacksquare$
	\section{}\label{appenc2}
	Here we will prove Theorem.\ref{thec2}.\\
	To prove that $\mathcal{N}_{U}$ as not $k$-use non $n$-locality breaking, it is enough to find an input of $n$ two-qubit states $\varrho_1,..,\varrho_n$ such that the overall state $\varrho_{in}^{'}$ satisfies Eq.(\ref{up11}).\\
	Let us consider:
	\begin{eqnarray}
		\varrho_{i}&=&\frac{1}{2}(|00\rangle\langle 00|+|11\rangle\langle11|-|11\rangle\langle 00|\nonumber\\
		&&|00\rangle\langle 11|)
	\end{eqnarray}
	So, $E_{i1}$$=$$E_{i2}$$=$$E_{i1}$$=$$1(\forall i$$=$$1,2,...,n).$\\
	Let the channel $\mathcal{N}_{U},$ satisfying conditions laid in the theorem, be used $k$ times in $\mathbf{N}_{lin}.$\\
	As discussed in Appendix.\ref{appenc1}, considering all possible cases of applying $\mathcal{N}_{U}$ $k$ times in $\mathbf{N}_{lin},$ $\textbf{W}_i^{'}$ are given by Eqs.(\ref{appa5},\ref{appa6},\ref{appa7}).\\
	$\forall i$$=$$1,2,...,n$ $E_{i1}^{'}$$\geq$$E_{i2}^{'}$$\geq$$E_{i3}^{'}$ denote the ordered singular values of $\textbf{W}_i^{'}.$ So for $j$$=$$1,2,3,$ $E_{ij}^{'}$$\in$$\{w_{i1}^{'},w_{i2}^{'},w_{i3}^{'}\}$ \\
	The conditions over $\alpha,\beta$ given in the theorem lead to the following possible cases:
	\begin{enumerate}
		\item [i] $(Re(\alpha),Re(\beta))$$=$$(0,0)$
		\item [ii] $(Im(\alpha),Im(\beta))$$=$$(0,0)$
		\item [iii] $(Re(\alpha),Im(\beta))$$=$$(0,0)$
		\item [iv] $(Im(\alpha),Re(\beta))$$=$$(0,0)$
		\item [v] $(Re(\alpha),Im(\alpha))$$=$$(0,0)$
		\item [vi] $(Re(\beta),Im(\beta))$$=$$(0,0)$
	\end{enumerate}
	We will prove the theorem for case.(i). Similarly it can be proved for other five cases.\\
	\textbf{Case.(i):} Eq.(\ref{appa8}) gets simplified as follows:
	\begin{eqnarray}\label{appb8}
		s_1&=&  |-(Im(\alpha))^2+(Im(\beta))^2|\nonumber\\
		s_2&=& |(Im(\alpha))^2+(Im(\beta))^2|\nonumber\\
		s_3&=& |(Im(\alpha))^2-(Im(\beta))^2|\nonumber\\
		&&
	\end{eqnarray}
	Now, as conditions imposed over parameters of any unital channel, by Eq.(\ref{ch0}), we have :
	\begin{equation}\label{ch00}
		Im(\alpha)^2+Im(\beta)^2=1
	\end{equation}
	Using Eq.(\ref{ch00}), from Eq.(\ref{appb8}), we get:
	\begin{eqnarray}\label{appb9}
		s_1&=& s_3= |-(Im(\alpha))^2+(Im(\beta))^2|\nonumber\\
		s_2&=& 1 \nonumber\\
		&&
	\end{eqnarray}
	Using Eq.(\ref{appb9}) and forms of $\textbf{W}_i^{'}$, given by Eqs.(\ref{appa5},\ref{appa6},\ref{appa7}), we get explicit forms of $E_{ij}^{'}:$
	\begin{eqnarray}\label{appb10}
		E_{i1}^{'}&=&1; \,\,	E_{i2}^{'}=E_{i3}^{'}= |-(Im(\alpha))^2+(Im(\beta))^2|\nonumber\\
		&&\forall i=1,2,...,m_1\nonumber\\
		E_{i1}^{'}&=&1; \,\,	E_{i2}^{'}=E_{i3}^{'}= |-(Im(\alpha))^2+(Im(\beta))^2|^2\nonumber\\
		&&\forall i=m_1+1,m_1+2,...,m_1+m_2\nonumber\\
		E_{i1}^{'}&=&E_{i2}^{'}=E_{i3}^{'}=1\,\,
		\forall i=m_1+m_2+1,m_1+m_2+2,...,n\nonumber\\
	\end{eqnarray}
	Upper bound($\textbf{B}_{n-lin}$) of Eq.(\ref{ineqb}) is given by:
	\begin{eqnarray}\label{appb11}
		\textbf{B}_{n-lin}&=&\sqrt{\Pi_{i=1}^n E_{i1}^{'}+\Pi_{i=1}^n E_{i2}^{'}}\nonumber\\
		&=&\sqrt{1+|-(Im(\alpha))^2+(Im(\beta))^2|^{m_1+2m_2}}\nonumber\\
		&=&\sqrt{1+|-(Im(\alpha))^2+(Im(\beta))^2|^{k}}\,\,\textmd{\small{By Eq.(\ref{k-val})}}\nonumber\\
	\end{eqnarray}
	Clearly $\textbf{B}_{n-lin}$$>$$1.$ So Eq.(\ref{up11}) is satisfied. This in turn implies detection of non $n$-local correlations via violation of $n$-local inequality(Eq.(\ref{ineqb})) in $\mathbf{N}_{lin}.$ Hence,$\mathcal{N_{U}}$ with $(Re(\alpha),Re(\beta))$$=$$(0,0)$ cannot act as a $k$-use non $n$-locality breaking channel in $\mathbf{N}_{lin}.$\\
	Similarly we can prove for other five cases.\\
	This completes the proof of the theorem.$\blacksquare$
	\section{}\label{appenc3}
	Here we will give proof of Theorem.\ref{thec3}.
\subsection{Representing $\varrho_i$ in Pauli Basis}
$\forall i$$=$$1,2,...,n, $ let $\varrho_i$ be any two-qubit state. Recall that in Bloch matrix representation(Eq.(\ref{st41})), $\varrho_i$ can be written as:
\begin{equation}\label{appc1}
	\small{\varrho}_i=\small{\frac{1}{4} (\mathbb{I}_{2} \times \mathbb{I}_{2}+\vec{a}_i.\vec{\sigma}\otimes \mathbb{I}_2+\mathbb{I}_2\otimes \vec{b}_i.\vec{\sigma}+\sum_{j=1}^{3}w_{i,j}\sigma_{j}\otimes\sigma_{j})},
\end{equation}
In Eq.(\ref{appc1}), $\vec{a}_i$$=$$(a_{i,1},a_{i,2},a_{i,3}),$ $\vec{b}_i$$=$$(b_{i,1},b_{i,2},b_{i,3}),$ with $||\vec{a}_i||,||\vec{b}_i||$$\leq$$1.$\\
Let $B_P$ denote two dimensional Pauli basis:
\begin{eqnarray}\label{appc2}
	B_P&=&\{\sigma_i\}_{i=0}^3.
\end{eqnarray}
Let $\{\vec{\delta}_j\}_{j=0}^3$ denote the basis vectors with respect to the basis $B_P$(Eq.(\ref{appc2})):
\begin{eqnarray}\label{appc3}
	\vec{\delta}_0&=&	\left[ {\begin{array}{c}
			1\\
			0\\
			0\\
			0\\		
	\end{array} } \right]_{B_P}\nonumber\\
	\vec{\delta}_1&=&	\left[ {\begin{array}{c}
			0\\
			1\\
			0\\
			0\\		
	\end{array} } \right]_{B_P}\nonumber\\
	\vec{\delta}_2&=&	\left[ {\begin{array}{c}
			0\\
			0\\
			1\\
			0\\		
	\end{array} } \right]_{B_P}\nonumber\\
	\vec{\delta}_3&=&	\left[ {\begin{array}{c}
			0\\
			0\\
			0\\
			1\\		
	\end{array} } \right]_{B_P}
\end{eqnarray}
Using these basis vectors(Eq.(\ref{appc3})), $\forall i=1,2,...,n$ $\varrho_i$ can be written as:
\begin{eqnarray}\label{appc4}
	\varrho_i&=&\sum_{j_1,j_2=0}^3\gamma_{i,j_1,j_2}\vec{\delta}_{j_1}\otimes \vec{\delta}_{j_2}\,\,\textmd{where}\nonumber\\
	\gamma_{i,0,0}&=&\frac{1}{4};\,\,\gamma_{i,0,j}=\frac{b_{i,j}}{4}\,\,j=1,2,3\nonumber\\
	\gamma_{i,j,0}=\frac{a_{i,j}}{4};\,\,\,\gamma_{i,j,j}&=&\frac{w_{i,j}}{4};\,\,j=1,2,3\nonumber\\
	\gamma_{i,j_1,j_2}&=&0\,\,\,j_1,j_2=1,2,3\,\,\,j_1\neq j_2
\end{eqnarray}
\subsection*{Effect of a Class of Non-Unital Channel $\mathcal{N}_{NU}$(Eqs.(\ref{nu1}),(\ref{nu2}))}
Under action of $\mathcal{N}_{NU}$(Eqs.(\ref{nu1},\ref{nu2})) the basis vectors change as follows:
\begin{eqnarray}\label{appc5}
	\vec{\delta}_0^{\tiny{(new)}}=\mathcal{N}_{NU}(\vec{\delta}_0)&=&	\left[ {\begin{array}{c}
			1\\
			0\\
			0\\
			t\\		
	\end{array} } \right]_{B_P}\nonumber\\
	\vec{\delta}_1^{\tiny{(new)}}=\mathcal{N}_{NU}(\vec{\delta}_1)&=&	\left[ {\begin{array}{c}
			0\\
			\lambda_1\\
			0\\
			0\\		
	\end{array} } \right]_{B_P}\nonumber\\
	\vec{\delta}_2^{\tiny{(new)}}=\mathcal{N}_{NU}(\vec{\delta}_2)&=&	\left[ {\begin{array}{c}
			0\\
			0\\
			0\\
			0\\		
	\end{array} } \right]_{B_P}\nonumber\\
	\vec{\delta}_3^{\tiny{(new)}}=\mathcal{N}_{NU}(\vec{\delta}_3)&=&	\left[ {\begin{array}{c}
			0\\
			0\\
			0\\
			\lambda_3\\		
	\end{array} } \right]_{B_P}
\end{eqnarray}
These transformed vectors will now be used to write the transformed state $\varrho_i^{'}$ after being acted upon by $\mathcal{N}_{NU}.$ For that we consider following two cases:
\begin{enumerate}
	\item single qubit of $\varrho_i$
	\item both qubits of $\varrho_i.$
\end{enumerate}
Case.$1:$ W.L.O.G., let us fix $i$$=$$1$ and let $\mathcal{N}_{NU}$ act on first qubit of $\varrho_1:$
\begin{eqnarray}\label{appc6}
	\varrho_1^{'}&=&\mathcal{N}_{NU}\otimes id_2(\varrho_{1})\nonumber\\
	&=&\mathcal{N}_{NU}\otimes id_2(\sum_{j_1,j_2=0}^3\gamma_{1,j_1,j_2}\vec{\delta}_{j_1}\otimes \vec{\delta}_{j_2})\nonumber\\
	&=&\sum_{j_1,j_2=0}^3\gamma_{1,j_1,j_2}\mathcal{N}_{NU}(\vec{\delta}_{j_1})\otimes \vec{\delta}_{j_2}\nonumber\\
	&=&\sum_{j_1,j_2=0}^3\gamma_{1,j_1,j_2}\vec{\delta}^{(\tiny{new})}_{j_1}\otimes \vec{\delta}_{j_2}\nonumber\\
	&=& \sum_{j_1,j_2=0}^3\gamma_{1,j_1,j_2}^{'}\vec{\delta}_{j_1}\otimes \vec{\delta}_{j_2}\,\textmd{where}\nonumber\\
	\gamma^{'}_{1,0,j}&=&\gamma_{1,0,j}\,j=0,1,2,3\nonumber\\
	\gamma^{'}_{1,3,0}&=&t \gamma_{1,0,0}+\lambda_3\cdot \gamma_{1,3,0}\nonumber\\
	\gamma^{'}_{1,3,j}&=&t\cdot \gamma_{1,0,j}\,\,j=1,2,3\nonumber\\
	\gamma^{'}_{1,1,0}&=&\lambda_1\cdot \gamma_{1,1,0}\nonumber\\
	\gamma^{'}_{1,j,j}&=&\lambda_j\cdot \gamma_{1,j,j}\,\,j=1,3\nonumber\\
\end{eqnarray}
As per our requirement, we need to focus particularly on the change in the correlation tensor $\mathbf{W}_1$$\Rightarrow$$\mathbf{W}_1^{'}.$ The correlation tensor($\mathbf{W}_1^{'}$) of the transformed state $\varrho_1^{'}$ is given by:
\begin{eqnarray}\label{appc7}
	\mathbf{W}_1^{'}&=&	\left[ {\begin{array}{ccc}
			\lambda_1\cdot w_{1,1}	&0&0\\
			0&0&0\\
			t\cdot b_{1,1}&t\cdot b_{1,2}&t\cdot b_{1,3}+\lambda_3\cdot w_{1,3}\\		
	\end{array} }\right]
\end{eqnarray}

Case.$2:$ W.L.O.G., let us fix $i$$=$$1$. $\mathcal{N}_{NU}$ acts on both qubits of $\varrho_1.$
\begin{eqnarray}\label{appc8}
	\varrho_1^{'}&=&\mathcal{N}_{NU}\otimes \mathcal{N}_{NU}(\varrho_{1})\nonumber\\
	&=&\mathcal{N}_{NU}\otimes \mathcal{N}_{NU}(\sum_{j_1,j_2=0}^3\gamma_{1,j_1,j_2}\vec{\delta}_{j_1}\otimes \vec{\delta}_{j_2})\nonumber\\
	&=&\sum_{j_1,j_2=0}^3\gamma_{1,j_1,j_2}\vec{\delta}^{(\tiny{new})}_{j_1}\otimes \vec{\delta}^{(\tiny{new})}_{j_2}\nonumber\\
\end{eqnarray}
$\varrho_1^{'}$ can be further simplified as we did in Case.$1.$\\
Fetching the correlation tensor($\mathbf{W}_1^{'}$) of the transformed state $	\varrho_1^{'}$ in this case, we get:
\begin{eqnarray}\label{appc9}
	\mathbf{W}_1^{'}&=&	\left[ {\begin{array}{ccc}
			\lambda_1^2\cdot w_{1,1}	&0&a_{1,1}\cdot \lambda_1\cdot t\\
			0&0&0\\
			b_{1,1}\cdot \lambda_1\cdot t&0&S_1\\		
	\end{array} }\right]\,\,\textmd{\small{where}}\\
	S_1&=&t^2+t\cdot\lambda_3(b_{1,3}+a_{1,3}) +\lambda_3^2\cdot w_{1,3}\nonumber
\end{eqnarray}
Now that we have seen how correlation tensor($\textbf{W}$) of any two-qubit state $\varrho_i$ changes when any non-unital channel $\mathcal{N}_{NU}$ from the class prescribed by Eq.(\ref{nu2}) acts on single or both qubits of $\varrho_i,$ we next use the transformed correlation tensors to prove our theorem.
\subsection{Proof of Theorem.\ref{thec3}}
Here we will prove Theorem.\ref{thec3}. For that we consider as before the most general situation that encompasses all possible cases of action of $\mathcal{N}_{NU}$ over some or all qubits distributed in network.\\
W.L.O.G., let $\mathcal{N}_{NU}$ act upon single qubit of each $\varrho_1,\varrho_2,...,\varrho_{m_1}$ and both qubits of each of $\varrho_{m_1+1},\varrho_{m_1+2},...,\varrho_{m_1+m_2}$ whereas does not act upon any qubit of remaining $n$$-$$m_1$$-$$m_2$ states $\varrho_{m_1+m_2+1},$$\varrho_{m_1+m_2+2},$...$\varrho_{n}.$
The overall state in the network $\varrho_{in}^{'}$ is given by Eq.(\ref{nb4}), i.e.,
\begin{eqnarray*}
	\varrho_{in}^{'}&=&\otimes_{j=1}^{m_1}(\mathcal{N}_{NU}\otimes id(\varrho_j))\otimes_{j=m_1+1}^{m_1+m_2}(\mathcal{N}_{NU}\nonumber\\
	&&\otimes \mathcal{N}_{NU}(\varrho_j))\otimes_{j=m_1+m_2+1}^n\varrho_j
\end{eqnarray*}
$\forall i$$=$$1,2,...,m_1,$ Eq.(\ref{appc7}) provides the correlation tensor $\textbf{W}_i^{'}$ of $\varrho_i^{'}:$

\begin{eqnarray}\label{appc10}
	\mathbf{W}_i^{'}&=&	\left[ {\begin{array}{ccc}
			\lambda_1\cdot w_{i,1}	&0&0\\
			0&0&0\\
			t\cdot b_{i,1}&t\cdot b_{i,2}&t\cdot b_{i,3}+\lambda_3\cdot w_{i,3}\\		
	\end{array} }\right]
\end{eqnarray}
$\forall i$$=$$1,2,...,m_1,$ ordered eigen values of $\mathbf{W}_i^{'}$  $E_{i1}^{'}$$\geq$$E_{i2}^{'}$$\geq$$E_{i3}^{'}$ are given by:
\begin{eqnarray}\label{appc100}
	E_{ij}^{'}&=&\frac{f_{i,1}+(-1)^j \sqrt{f_{i,2}}}{2};\,\,\,j=1,2;\,\, 	E_{i3}^{'}=0,\,\,\textmd{\small{where}}\nonumber\\
	f_{i,1}&=&||\vec{b}||^2\cdot t+\sum_{k=1,3}\lambda_k^2\cdot w_{i,k}^2+2 t \cdot b_{i,3}\cdot w_{i,3}\cdot \lambda_3\nonumber\\
	f_{i,2}&=& f_{i,1}^2-4((b_{i,2}\cdot w_{i,1})^2\cdot t\cdot \lambda_1)^2\nonumber\\
	&+&(b_{i,3}\cdot  w_{i,1}\cdot t\cdot \lambda_1)^2+(\lambda_1\cdot \lambda_3\cdot w_{i,1}\cdot w_{i,3})^2\nonumber\\
	&+&2( w_{i,1}\cdot \lambda_1)^2b_{i,3}\cdot w_{i,3}\cdot t\cdot \lambda_3)\nonumber\\
	&=&f_{i,1}^2-4((b_{i,2}\cdot w_{i,1}\cdot t\cdot \lambda_1)^2\nonumber\\
	&+&(w_{i,1}\cdot \lambda_1)^2(b_{i,3}\cdot t+w_{i,3}\cdot \lambda_3)^2)\nonumber\\
	&&\quad\quad \forall \,i=1,2,...,m_1;\nonumber\\
	&&
\end{eqnarray}
It is clear from Eq.(\ref{appc100}) that $\sqrt{f_{i,2}}$$\leq$$f_{i,1}.$ Hence $E_{i,1}^{'},E_{i,2}^{'},E_{i,3}^{'}$$\geq$$0$ $\forall i$$=$$1,2,...,m_1.$\\
Also, we get:
\begin{eqnarray}\label{appc10o}
	E_{i,1}^{'}	+E_{i,2}^{'}&=&f_{i,1}\nonumber\\
	&\leq& |t|+\sum_{k=1,3}\lambda_k^2+2|t\cdot \lambda_3|\,\,\textmd{\small{as }}||b||,w_{i,k},b_{i,3}\leq 1\nonumber\\
	&=&(|t|+|\lambda_3|)^2+\lambda_1^2\,\,\forall i= 1,2,...,m_1.\nonumber\\
	&&
\end{eqnarray}
Now 
Using Eq.(\ref{appc9}), $\forall i$$=$$m_1$$+$$1,m_1$$+$$2,$...,$m_1$$+$$m_2$ the correlation tensor $\textbf{W}_i^{'}$ of $\varrho_i^{'}$ is given by:
\begin{eqnarray}\label{appc11}
	\mathbf{W}_i^{'}&=&	\left[ {\begin{array}{ccc}
			\lambda_1^2\cdot w_{i,1}	&0&a_{i,1}\cdot \lambda_1\cdot t\\
			0&0&0\\
			b_{i,1}\cdot \lambda_1\cdot t&0&S_i\\		
	\end{array} }\right]\,\,\textmd{\small{where}}\\
	S_i&=&t^2+t\cdot\lambda_3(b_{i,3}+a_{i,3}) +\lambda_3^2\cdot w_{i,3}\nonumber
\end{eqnarray}
$\forall i$$=$$m_1$$+$$1,$...,$m_1$$+$$m_2$ ordered eigen values of $\mathbf{W}_i^{'}$  $E_{i1}^{'}$$\geq$$E_{i2}^{'}$$\geq$$E_{i3}^{'}$ are given by:
\begin{eqnarray}\label{appc12}
	E_{ij}^{'}&=&\frac{f_{i,3}+(-1)^j \sqrt{f_{i,4}}}{2};\,\,\,j=1,2;\,\, 	E_{i3}^{'}=0,\,\,\textmd{\small{where}}\nonumber\\
	f_{i,3}&=&(b_{i,1}^2+a_{i,1}^2)t^2\cdot \lambda_1^2+w_{i,1}^2\lambda_1^4+S_i^2\nonumber\\
	f_{i,4}	&=&f_{i,3}^2-4\lambda_1^4(S_i\cdot w_{i,1}-a_{i,1}\cdot b_{i,1}\cdot t^2)^2\nonumber\\
	&&\quad\,\, \forall \,i=m_1+1,m_1+2,...,m_1+m_2;\nonumber\\
	&&
\end{eqnarray}
Here also, we have $E_{i,1}^{'},E_{i,2}^{'},E_{i,3}^{'}$$\geq$$0$ $\forall i$$=$$m_1+1,...,m_1+m_2$ and
\begin{eqnarray}\label{appc10oo}
	E_{i,1}^{'}	+E_{i,2}^{'}&=&f_{i,3}\nonumber\\
	&<&	2t^2\cdot \lambda_1^2+\lambda_1^4+\nonumber\\
	&&(t^2+|t|\cdot |b_{i,3}+a_{i,3}|\cdot |\lambda_3|+\lambda_3^2\cdot |w_{i,3}|)^2\nonumber\\
	&<&2t^2\cdot \lambda_1^2+\lambda_1^4+\nonumber\\
	&&(|t|+|\lambda_3|)^4\,\,\forall i= m_1+1,...,m_1+m_2.
\end{eqnarray}
$\forall i$$=$$m_1$$+$$m_2$$+$$1,$...$n,$  correlation tensor $\textbf{W}_i^{'}$ of $\varrho_i$ remains unaltered:
\begin{eqnarray}\label{appc13}
	\textbf{W}_i^{'}&=&\textmd{diag}(w_{i1}^{'},w_{i2}^{'},w_{i3}^{})\,\,\textmd{\small{where}}\nonumber\\
	w^{'}_{i1}&=&w_{i1}\nonumber\\
	w^{'}_{i2}&=&w_{i2}\nonumber\\
	w^{'}_{i3}&=& w_{i3}\,\forall i=m_1+m_2+1,...,n\nonumber\\
	&&
\end{eqnarray}
Hence $\forall i$$=$$m_1$$+$$m_2$$+$$1,$...$n,$ $E_{i,j}^{'}$$=$$E_{i,j}.$

Upper bound($\textbf{B}_{n-lin}$) of Eq.(\ref{ineqb}) is given by:
\begin{eqnarray}\label{appc14}
	\textbf{B}_{n-lin}&=&\sqrt{\Pi_{i=1}^n E_{i1}^{'}+\Pi_{i=1}^n E_{i2}^{'}}\nonumber\\
	&\leq& \sqrt{\Pi_{i=1}^{m_1+m_2} E_{i1}^{'}+\Pi_{i=1}^{m_1+m_2} E_{i2}^{'}}\,\textmd{\small{as all }}E_{i,j}\leq 1\nonumber\\
	&\leq&\sqrt{\Pi_{i=1}^{m_1+m_2}(E_{i1}^{'}+E_{i2}^{'})},\,\textmd{\small{as all }}E_{ij}^{'}\geq 0\nonumber\\
	&=&\sqrt{\Pi_{i=1}^{m_1}f_{i,1}\Pi_{i=m_1+1}^{m_1+m_2}f_{i,3}}\,\,\textmd{\small{By Eqs.(\ref{appc10o},\ref{appc10oo})}}.\nonumber\\
	&<&((|t|+|\lambda_3|)^2+\lambda_1^2)^{\frac{m_1}{2}}\nonumber\\
	&&(2t^2\cdot \lambda_1^2+\lambda_1^4+	(|t|+|\lambda_3|)^4)^{\frac{m_2}{2}}\nonumber\\
	&<&1\,\, \textmd{\small{By given conditions Eq.(\ref{resnon1}) over }}\mathcal{N}_{NU}
\end{eqnarray}
This implies that $\varrho_{in}^{'}$ is not detected as a non $n$-local resource by the $n$-local inequality. \\
$\varrho_1,\varrho_2,...,\varrho_n$ being arbitrary, $\mathcal{N}_{NU}$ turns out to be $k$-use non $n$-locality breaking channel.
This proves the theorem.$\blacksquare$
\section{}\label{appenc4}
The steps of the proof are same as that used in proof of Theorem.\ref{thec3}. All the expressions involving elements, singular values $E_{ij}^{'}$  of $\textbf{W}_i^{'}(i$$=$$1,2,...,n)$ remain the same. \\
We thus use $E_{ij}^{'}$ and related expressions provided in Eqs.(\ref{appc100},\ref{appc11},\ref{appc12},\ref{appc13},\ref{appc14}) to work with the upper bound $\mathbf{B}_{n-star}$(Eq.(\ref{boundstar})) of $n$-local inequality Eq.(\ref{ineqs}). \\
$\mathbf{B}_{n-star}$ thus takes the form:
\begin{eqnarray}\label{appdr14}
	\textbf{B}_{n-star}&=&\sqrt{\Pi_{i=1}^n (E_{i1}^{'})^{\frac{2}{n}}+\Pi_{i=1}^n (E_{i2}^{'}})^{\frac{2}{n}}\nonumber\\
	&\leq&\sqrt{\Pi_{i=1}^{m_1+m_2} (E_{i1}^{'})^{\frac{2}{n}}+\Pi_{i=1}^{m_1+m_2} (E_{i2}^{'}})^{\frac{2}{n}}\nonumber\\
	&&	\textmd{\small{as }}E_{ij}^{'}\leq 1\,\forall i=m_1+m_2+1,...,n\nonumber\\
	&\leq& \sqrt{\Pi_{i=1}^{m_1+m_2} ((E_{i1}^{'})^{\frac{2}{n}}+ (E_{i2}^{'})^{\frac{2}{n}})}\,\,\textmd{\small{as all }}E_{ij}^{'}\geq 0\nonumber\\
	&=& \sqrt{\Pi_{i=1}^{m_1}\sum_{j=1}^2(\frac{f_{i,1}+(-1)^j \sqrt{f_{i,2}}}{2})^{\frac{2}{n}}}\cdot\nonumber\\
	&&\sqrt{\Pi_{i=m_1+1}^{m_1+m_2}\sum_{j=1}^2(\frac{f_{i,3}+(-1)^j \sqrt{f_{i,4}}}{2})^{\frac{2}{n}}}\nonumber\\
	&\leq&\sqrt{2^{m_1}\Pi_{i=1}^{m_1}(\frac{f_{i,1}}{2})^{\frac{2}{n}}\cdot 2^{m_2}\Pi_{i=m_1+1}^{m_1+m_2}(\frac{f_{i,3}}{2})^{\frac{2}{n}}}\,(\textmd{\small{Using Jensen's inequality over concave function }}g(t)=t^{\frac{2}{n}},\,n\geq 2) \nonumber\\
	&\leq& \sqrt{2^{(m_1+m_2)}}\sqrt{\Pi_{i=1}^{m_1}(\frac{(|t|+|\lambda_3|)^2+\lambda_1^2}{2})^{\frac{2}{n}}}\cdot\nonumber\\
	&&\sqrt{\Pi_{i=m_1+1}^{m_1+m_2}(\frac{2t^2\cdot \lambda_1^2+\lambda_1^4+(|t|+|\lambda_3|)^4}{2})^{\frac{2}{n}}}\,\,\textmd{\small{By Eqs.(\ref{appc10o},\ref{appc10oo})}}\nonumber\\
	&=&\sqrt{2^{(m_1+m_2)(1-\frac{2}{n})}(2t^2\cdot \lambda_1^2+\lambda_1^4+(|t|+|\lambda_3|)^4)^{\frac{2\cdot m_2}{n}}}\cdot \sqrt{((|t|+|\lambda_3|)^2+\lambda_1^2)^{\frac{2\cdot m_1}{n}}}\nonumber\\
	&=&2^{(m_1+m_2)\frac{n-2}{2n}(2t^2\cdot \lambda_1^2+\lambda_1^4+(|t|+|\lambda_3|)^4)^{\frac{m_2}{n}}}\cdot \nonumber\\
	&&((|t|+|\lambda_3|)^2+\lambda_1^2)^{\frac{m_1}{n}}\nonumber\\
	&<&1\,\,\textmd{\small{By given conditions Eq.(\ref{resnon2}) over }}\mathcal{N}_{NU}\nonumber\\
	&&
\end{eqnarray}
\section{}\label{appenc5}
Here we are dealing with class of non-unital channels. So, to prove this theorem we use all the possible cases of correlation tensors as used in proof of theorem.\ref{thec3}. \\
To prove theorem.\ref{thec7} for any $(n,m_1,m_2)$, it is sufficient to find at least one collection of $n$ number of two-qubits states such that when these states are used in $\mathbf{N}_{star}, $ any non-unital channel satisfying Eq.(\ref{resnon3}) will not destroy non $n$-local correlations when used in a pattern(as per $m_1,m_2$) as prescribed in the theorem.\\
$\forall i$$=$$1,2,...,n$, let $\varrho_i$$=$$\frac{|00\rangle\langle 00|+|11\rangle\langle 11|+|00\rangle\langle 11|+|11\rangle\langle 00|}{2}.$ Correlation tensor of this state is $\textmd{diag}(1,-1,1).$ So $w_{i,1}$$=$$w_{i,3}$$=$$1$ and $w_{i,2}$$=$$-$$1.$ Also it has no local Bloch vector. So $a_{i,j}$$=$$b_{i,j}$$=$$0, \forall j$$=$$1,2,3.$\\
When non-unital channel $\mathcal{N}_{NU}$ is used on single qubit(say first one) of any of $\varrho_i$ then correlation tensor($\mathbf{W}_i^{'}$) of $\varrho_i^{'}$ is given by Eq.(\ref{appc7}):
\begin{eqnarray}\label{appcs7}
	\mathbf{W}_i^{'}&=&	\left[ {\begin{array}{ccc}
			\lambda_1	&0&0\\
			0&0&0\\
			0&0&\lambda_3\\		
	\end{array} }\right]
\end{eqnarray}
When non-unital channel $\mathcal{N}_{NU}$ is used on both qubits of any of $\varrho_i$ then correlation tensor of $\varrho_i^{'}$ is given by Eq.(\ref{appcs9}):
\begin{eqnarray}\label{appcs9}
	\mathbf{W}_i^{'}&=&	\left[ {\begin{array}{ccc}
			\lambda_1^2	&0&0\\
			0&0&0\\
			0&0&t^2+\lambda_3^2 \\		
	\end{array} }\right]
\end{eqnarray}
Using $W_1,W_2,....W_{m_1}$ as given by Eq.(\ref{appcs7}), $W_{m_1+1},W_{m_1+2},....W_{m_1+m_2}$ as given by Eq.(\ref{appcs9}) and correlation tensors of remaining $n-m_1$$-$$m_2$ states as $\textmd(diag)(1,-1,1)$(unchanged), the upper bound of Eq.(\ref{ineqs}) gives:
\begin{eqnarray}\label{appd14}
	\textbf{B}_{n-star}&=&\sqrt{\Pi_{i=1}^n (E_{i1}^{'})^{\frac{2}{n}}+\Pi_{i=1}^n (E_{i2}^{'}})^{\frac{2}{n}}\nonumber\\
	&=&\sqrt{Max[C_1,C_2]^{\frac{m_1}{n}}\cdot Max[C_3,C_4]^{\frac{m_2}{n}}+Min[C_1,C_2]^{\frac{m_1}{n}}\cdot Min[C_3,C_4]^{\frac{m_2}{n}}}\nonumber\\
	\textmd{\small{with }}&& C_1=\lambda_1^2;\, C_2=\lambda_3^2 \nonumber\\
	C_3&=&\lambda_1^4;\,C_4=(t^2+\lambda_3^2)^2\nonumber\\
	&>&1\, \textmd{\small{By Eq.(\ref{resnon3})}}
\end{eqnarray}
This proves the theorem.
	
	\section*{Data Availability}
	No data were created or analyzed in this study.
	
\end{document}